\documentclass[superscriptaddress,twocolumn]{revtex4}

\usepackage{amsmath,amssymb,graphicx,color,amsfonts,hyperref}
\usepackage{framed}

\begin{document}	
\title{Effect of weak nonlocal nonlinearity on generalized sixth-order dispersion modulational instability in optical media}

\author{Conrad B. Tabi}
	\email {tabic@biust.ac.bw}
	\affiliation{Department of Physics and Astronomy, Botswana International University of Science and Technology, Private Mail Bag 16 Palapye, Botswana }
	
	\author {Camus G. Latchio Tiofack}
\email {glatchio@yahoo.fr}
\affiliation{Faculty of Sciences, University of Maroua, P.O. Box 814, Maroua, Cameroon}

\author{Hippolyte Tagwo}
	\affiliation{University Institute of Wood Technology, P.O. Box 306, Mbalmayo, Cameroon}
	\affiliation{Laboratory of Mechanics, Department of Physics, Faculty of Science, University of Yaound\'{e} I, P.O. Box 812, Yaound\'{e}, Cameroon}
		\email{htagwo@gmail.com}

\author{Timol\'eon  C. Kofan\'e}
\email {tckofane@yahoo.com}
	\affiliation{Department of Physics and Astronomy, Botswana International University of Science and Technology, Private Mail Bag 16 Palapye, Botswana }
	\affiliation    {Laboratory of Mechanics, Department of Physics, Faculty of Science, University of Yaound\'{e} I,
P.O. Box 812, Yaound\'{e}, Cameroon}
	\date{\today}

\begin{abstract}
This paper analyzes the behaviors of solitons in even higher-order dispersive media and explores the modulational instability phenomenon in optical media. The analysis considers quadratic, quartic, and sextic dispersions with weakly nonlocal Kerr nonlinearity. The results show that nonlocality enhances the MI gain and leads to rogue waves in response to different combinations of even dispersions and nonlocal nonlinearity. The study suggests that Kerr nonlocality can enhance the excitation of extreme events in even higher-order dispersive nonlinear media with potential applications in optical fibers and fiber lasers.
\end{abstract}
\maketitle

\section{Introduction}\label{sec:intro}

Nonlinear dynamics of electromagnetic waves inside a medium is one of the sources of integrable, near-integrable, and non-integrable nonlinear evolution equations of mathematical physics having solutions of solitonic and solitonic dissipative types, and which is connected with plenty of interesting and practically important phenomena. It is well-known that the self-phase modulation of waves propagating in nonlinear optical fiber is governed by the one-dimensional cubic nonlinear Schr\"odinger (NLS) equation which is an integrable partial differential equation possessing an infinite number of symmetries corresponding to an infinite number of conserved quantities.  In the case of temporal solitons, the balance between the self-focusing nonlinearity (negative Kerr-nonlinearity) and the anomalous dispersion regime (positive group velocity) leads to the bright solitons, while the interplay between the self-defocusing nonlinearity (positive Kerr-nonlinearity) and the normal dispersion regime (negative group velocity), leads to the dark solitons. In the case of spatial solitons, the balance between Kerr-nonlinearity and diffraction can exhibit bright or dark solitons for the self-focusing and self-defocusing nonlinear media, respectively. Beside the temporal and spatial optical solitons, there exist optical structures which are fully confined in space and time and are referred to as spatiotemporal optical solitons, also known as "light bullets" ~\cite{LambSoliton1980,AblowitzInverse1981,ZakharovInverse1984,DoddSolitons1982,KivsharPhotonics2003,SilberbergOPlet1990}. The term {\it "optical bullet"} was proposed by Silberberg~\cite{SilberbergOPlet1990} to indicate a complete spatiotemporal soliton, for which confinement in the three spatial dimensions and localization in the temporal domain are achieved by the balance between a focusing nonlinearity and the spreading due to chromatic dispersion and angular diffraction.

Inverse scattering theory has been applied with great success to the NLS equation, and solutions are dark and bright solitons, with {\it sech} and {\it tanh} profiles, respectively. Under some physical conditions, the self-modulation of the carrier wave leads to a spontaneous energy localization via the generation of rational solitons which are localized in space and time, having envelope structures such as Akhmediev-Peregrine waves, Akhmediev wave trains, and Kuznetsov-Ma waves, respectively~\cite{KivsharPhotonics2003,KharifRW2009,SolliNature2007,PeregrineJAust1983,KuznetsovSovPhys1977,Ma1979,AkhmedievTMP1987}.

The possibility of using optical solitons not only as carriers of information in high-speed and long-haul communication systems but also in ultrashort lasers as well has been the focus of intense research activity for almost three decades, leading to various scenarios of the pulse evolution, which are resulted in particular changes of the pulse shape, spectrum, and chirp. Indeed, Optical solitons are very stable against changes in the properties of the medium, provided that these changes occur over distances that are long compared with a soliton period, and optical solitons can adapt their shape to slowly varying parameters of the medium. However, such optical solitons are perturbed by several higher-order dispersive and nonlinear effects, mainly when ultrashort optical pulses are used to excite them~\cite{AgrawalNFO1989}.

So said, for the specific case in which a fundamental soliton is perturbed by third-order dispersion (TOD) only~\cite{WaiPRA1990,ElginPRA1993,KarpmanPRE1993,KodamaOLett1994,AkhmedievPRA1995,ElginOpCom1995}, there is the emission of nonsolitonic radiation at a frequency different from those associated with the soliton~\cite{WaiPRA1990}, and analytic expressions for the frequency and amplitude of the nonsolitonic radiation peak that is generated in the output spectrum when the TOD perturbs a higher-order soliton have been found~\cite{RoyPRA2009}. Several studies suggest that effects of group-velocity dispersion (GVD) and fourth-order dispersion (FOD) can be minimized under some specific physical conditions, and new types of solitary wave solutions for second and FOD NLS equation only have been reported since the1990s~\cite{HookOpLett1993,AkhmedievOpCom1994,ChristovOpLett1994,KarlssonOptCom1994,AbdullaevOpCom1994,AkhmedievOpCom1995,PicheOpLett1996,ZakharovJETP1998,KalithasanJOpt2010,ArmaroliOpLett2014,RedondoNatCom2016,LoOpExp2018,TaheriOpLett2019}. In such cases, propagation of high-intensity ultrashort pulse can be accurately described by considering higher-order dispersion effects like FOD.  In particular, FOD effects lead the ultrashort solitons to decay due to dispersive radiation on optical-fiber-guided~\cite{HookOpLett1993}. In the mode-locked laser with the near-zero second as well as near-zero TOD, pulses much shorter than 10fs can be generated if FOD is further reduced~\cite{ChristovOpLett1994}.  An exact localized analytical solution has been obtained in the case where negative FOD is added to anomalous second-order dispersion~\cite{KarlssonOptCom1994}. In the case of an optical fiber near the zero dispersion point, a new region with modulational instability (MI) process of electromagnetic waves has been found, and a recurrence phenomenon similar to the Fermi-Pasta-Ulam problem has been observed for both positive and negative FOD~\cite{AbdullaevOpCom1994}. Moreover, it has been shown that the bright optical soliton's temporal shape and peak power can be stable when a weak TOD is introduced in the presence of FOD~\cite{PicheOpLett1996}. Indeed, higher-order dispersion effects in the anomalous dispersion regime can suppress the MI gain spectra of an electromagnetic wave in a resonant optical fiber with a two-level system~\cite{KalithasanJOpt2010}. Remarkably, the possibility of multiple instability peaks originating from the parametric resonance, coexisting with the conventional MI because of FOD has been thoroughly analyzed, where periodic variations of GVD in the proximity of the zero dispersion point have been considered~\cite{ArmaroliOpLett2014}.

The recently discovered pure-quartic solitons~\cite{RedondoNatCom2016,LoOpExp2018,TaheriOpLett2019,KruglovPRA2018,TamOpLett2019,GaoPRE2020,TamPRA2020,ParkerPhysD2021,BandaraPRA2021,ZhaoOPLett2021,ZhangOpLett2022,QianOpLett2022,YaoPRR2022,AlexanderOpLett2022,LiOpCom2022,QiangPRA2022,MartijnAPLPhot2021} arising from the interaction of quartic dispersion and Kerr nonlinearity, formed a framework for subsequent research in the field of communications, slow-light devices, and ultrafast laser science and open also the door to unexplored soliton regimes. A class of bright solitons arising purely from negative FOD and self-phase modulation interaction, which can occur even for normal group-velocity dispersion, has been demonstrated experimentally~\cite{RedondoNatCom2016}. Importantly, a general analysis of the dispersion and nonlinear properties necessary to observe pure-quartic solitons in optical platforms has been reported~\cite{LoOpExp2018}. Moreover, the Lagrangian variational method has been used to show that continuous wave coherent light of a Kerr resonator featuring quartic GVD forms a pure-quartic soliton with a Gaussian envelope~\cite{TaheriOpLett2019}. An exact solitary wave solution that applies for all orders of dispersion up to the fourth order has been proposed~\cite{KruglovPRA2018}. A family of pure-quartic solitons, existing through a balance of positive Kerr nonlinearity and negative quartic dispersion, has been found~\cite{TamOpLett2019}. The propagation velocities, periodicity, and localization of a Gaussian-type initial perturbation with cosine-type modulation on a plane wave inside a nonlinear fiber model purely with FOD have been predicted by the modified linear stability analysis~\cite{GaoPRE2020}. A combination of negative quartic and some quadratic dispersions leading to high soliton energies has been reported~\cite{TamPRA2020}. Multi-pulse solitary wave solutions' existence and spectral stability to an NLS with both second-order dispersion and FOD have been developed~\cite{ParkerPhysD2021}. A dynamical system approach has been successfully applied to the NLS equation in the presence of both quartic and quadratic dispersion terms, and families of solitons with different symmetry properties have been characterized~\cite{BandaraPRA2021}. The vector properties of quartic solitons in a pure FOD birefringent fiber and a quartic dispersion dominant mode-locked fiber laser have been investigated~\cite{ZhaoOPLett2021}. Indeed, the bifurcation diagrams in a passively mode-locked fiber laser have revealed that the pure-quartic solitons can alternate between the stable single soliton and pulsating regimes multiple times before transiting into the chaotic state~\cite{ZhangOpLett2022}. Moreover, it has been demonstrated that dissipative pure-quartic solitons in a fiber laser possesses a higher energy scaling ability compared to conventional dissipative solitons~\cite{QianOpLett2022}. The dynamic orbits outside the heteroclinic separatrix in the phase plane, corresponding to the Fermi-Pasta-Ulam recurrences and Akhmediev breathers (ABs), have been discussed in the pure quartic systems~\cite{YaoPRR2022}. Stable dark solitons in the presence of pure-quartic dispersion and both quadratic and quartic dispersion effects have been theoretically predicted~\cite{AlexanderOpLett2022}. The standard Gaussian trial solution has been modified to reflect the radiation-induced oscillation damping of the breathing pure-quartic solitons~\cite{LiOpCom2022}. Subsequently, bright solitons can exist, for the particular choice of the dispersion parameters, in the NLE equation with quadratic, quartic, and sextic dispersion effects~\cite{QiangPRA2022}.

Our primary objective in this paper is to investigate the MI process in a fiber system taking into account quadratic, quartic, and sextic dispersion terms as well as the inherently nonlocal character of nonlinearity~\cite{SalernoPRE2018,ZangaCNSNS2020}. Nonlocal nonlinearity means that the light-induced refractive index change of a material at a particular location is determined by the light intensity in a certain neighborhood of this location.  MI induced by the interplay between nonlinearity and dispersion (in the temporal domain) or diffraction (in the spatial domain) is a generic phenomenon in nonlinear physics. In particular, MI has been found to be relevant in understanding the formation and propagation of solitonic waves~\cite{KevrekidisPRA2003}, explaining the domain wall formation~\cite{ZhangPRL2005}, and quantum phase transition~\cite{SmerziPRL2002}. Our results suggest nonlocality to enhance MI even in some nonintegrable cases and additionally confirm MI to be a key mechanism for generating rogue waves (RW) under different combinations of even dispersions and nonlinear terms.

The remaining part of the paper is structured as follows. Section~\ref{s2} presents the governing weakly nonlocal model under study. Then, in Section~\ref{s3}, we investigate, through analytical methods, the MI excitation in view of the generation of bright solitary waves in the optical fiber. Section~\ref{s4} discusses the numerical experiment of the model equation. Section~\ref{s5} summarizes our investigations and concludes the work.

\section{Model}\label{s2}

The model proposed in the present work combines the effect of even-order dispersions up to the sixth order that interacts with nonlocal nonlinear effects for optical solitons to take place. This is modeled by the following modified NLS equation:  
\begin{equation}\label{eq1}
\begin{split}
i\frac{\partial \psi}{\partial z}&-\frac{\beta_2}{2}\frac{\partial^2\psi}{\partial t^2}+\frac{\beta_4}{24}\frac{\partial^4\psi}{\partial t^4}\\
&-\frac{\beta_6}{720}\frac{\partial^6\psi}{\partial t^6}+\Delta n\psi(z,t)=0,
\end{split}
\end{equation}
where $\psi(z,t)$ is the pulse complex envelope, $z$ is the propagation coordinate, and $t$ is the local time. The $\beta_n$ represent the $n^{th}-$order dispersion coefficients, with 
\begin{equation}
\beta_n=\frac{\partial^n \beta}{\partial\omega^n}=\frac{\partial^{n-1} v_g^{-1}}{\partial\omega^{n-1}},
\end{equation}
where $\beta$ is the mode propagation constant and $v_g$ is the group velocity. 

It is well-known in nonlinear Kerr media that the intensity-dependent refractive index change is given by $\Delta n(t,I)=I$, with the intensity of the beam $I=|\psi(z,t)|^2$. In the case of nonlinear media with a nonlocal nonlinearity, the nonlinear refractive index change of the medium can be represented by the following phenomenological model
\begin{equation}\label{eq2}
\begin{split}
\Delta n(t,I)&=\alpha\int^{+\infty}_{-\infty}R(t-t^{'})I(t',z)dt^{'},
\end{split}
\end{equation}
where $\alpha$ is the strength of the nonlocal Kerr nonlinearity. Under weak nonlocality, which is the main concern of this work, the above nonlinear term is expanded as 
\begin{equation}\label{eq4}
 \Delta n \left( t, I \right)=\alpha\left|\psi\left(t,z\right)\right|^{2}+\alpha\gamma\frac{\partial^2\left(\left|\psi\left(t,z\right)\right|^{2}\right)}{\partial t^2},
 \end{equation}
where $ \gamma=\frac{1}{2}\int _{-\infty }^{+\infty }t^{2}\!R\left(t\right){dt} $ is the nonlocal parameter. Inserting Eq.(\ref{eq4}) into Eq.(\ref{eq1}), one gets the following modified NLS equation:
\begin{equation}\label{eq5}
\begin{split}
i\frac{\partial \psi}{\partial z}&-\frac{\beta_2}{2}\frac{\partial^2\psi}{\partial t^2}+\frac{\beta_4}{24}\frac{\partial^4\psi}{\partial t^4}-\frac{\beta_6}{720}\frac{\partial^6\psi}{\partial t^6}\\
&+\alpha\left|\psi\right|^{2}\psi+\chi\frac{\partial^2\left(\left|\psi\right|^{2}\right)}{\partial t^2}\psi=0,
\end{split}
\end{equation}
where $\chi=\alpha\gamma$. Compared to the model used in Ref.~\cite{QiangPRA2022}, the nonlocality brings about the additional term $\chi\frac{\partial^2\left(\left|\psi\right|^{2}\right)}{\partial t^2}\psi$ to the equation, which can be switched off by imposing $\gamma=0$. In such a context, solitons with oscillating tails were numerically characterized along with their dynamical response to the interplay between higher-order dispersions and Kerr nonlinearity. Other studies have explored the cases $\gamma=0$, $\beta_2=\beta_6=0$, where the interplay between the negative quartic and Kerr nonlinearity was found to sustain the stability of pure-quartic solitons of the NLS equation~\cite{RungePRR2021}.  More recently, a modified linear stability analysis on the latter was proposed, where Gao et al.~\cite{GaoPRE2020} further proposed a modified gain of MI from a localized perturbation. Soliton properties can be modified due to the presence of  nonlocalities such as strong modification of modulational instability~\cite{KrolikowskiJOptB2004,TagwoOptik2017,ZangaCNSNS2020,TabiOpLett2022,TabiPRE2022}, suppression of beam collapse~\cite{BangPRE2002}, dramatic change of the soliton interaction~\cite{PecciantiOpLett2002,NikolovOpLett2004}, formation of multi-soliton bound states~\cite{XuOpLett2005}, stabilization of spatially localized vortex solitons~\cite{DesyatnikovPOpt2005}, symmetry breaking azimuthal instability~\cite{YakimenkoPRE2005,BriedisOpExp2005}  as well as stabilization of different nonlinear structures such as ring-like clusters of many solitons~\cite{DesyatnikovPRL2002} and modulated localized vortex beams or azimuthons~\cite{DesyatnikovPRL2005}. A simple procedure for estimating the strength of a weak nonlocality in the propagation of a Gaussian beam has been formulated, and the conditions for breathing soliton formation in both one and two transverse dimensions have been derived~\cite{BezuhanovPRA2008}. Moreover, the high limit of the nonlocal 2D NLS equation has been derived~\cite{SkupinPRE2006}. Indeed, the stabilization of 2D ring dark solitons and ring anti-dark solitons has been demonstrated in nonlocal media~\cite{ChenJPB2017}.

\section{Linear stability analysis and gain spectrum}\label{s3}

\subsection{Linear stability analysis}

To exclusively address the analytical study of MI in the above-introduced model, we assume it admits continuous wave (CW) solutions that propagates inside the fiber with the initial input power $P_{0}$ so that
\begin{equation}\label{eq6}
\psi(z,t) = \sqrt{P_{0}} e^{i\phi_{NL}z} ,
\end{equation}
where $\phi_{NL} = \alpha P_{0}$ is the nonlinear phase shift due to the nonlinearity. To further proceed, a slight perturbation is introduced into solution (\ref{eq6}) to study its stability, 
\begin{equation} \label{eq7}
\psi(z,t) = [\sqrt{P_{0}} + \zeta(z,t)]e^{i\phi_{NL}z} ,
\end{equation}
where $ \zeta(z,t)$ is a small perturbation, with $\left|\zeta(z,t)\right|\ll \sqrt{P_0} $ being the complex field. Substituting the perturbed solution (\ref{eq7}) into Eq.(\ref{eq5}) and linearizing around the unperturbed solution,  the perturbed field is found to be governed by the following equations: 
\begin{equation}\label{eq8}
\begin{split}
i\frac{\partial \zeta}{\partial z}&-\frac{\beta_2}{2}\frac{\partial^2\zeta}{\partial t^2}+\frac{\beta_4}{24}\frac{\partial^4\zeta}{\partial t^4}-\frac{\beta_6}{720}\frac{\partial^6\zeta}{\partial t^6}\\
&+\alpha P_0(\zeta+\zeta^*)+\chi P_0\left(\frac{\partial^2\zeta}{\partial t^2}+\frac{\partial^2\zeta^*}{\partial t^2}\right)=0,
\end{split}
\end{equation}
where $\zeta ^ \ast $ is the complex conjugate of the perturbed field.  Further, the solution for Eq.(\ref{eq7}) is taken in the form
\begin{align}\label{eq9}
\zeta(z,t) = \zeta_1(z)e^{-i\Omega t}  + \zeta_2(z) e^{i\Omega t}, 
\end{align}
where $ \zeta_1(z)$ and $\zeta_2(z)$ are the complex perturbation fields and $ \Omega$ is the complex modulation frequency. Applying the above ansatz to the perturbed field  Eq. (\ref{eq7})  leads to a $2 \times 2$ system of the following form  for the perturbation fields:
\begin{equation} \label{eq10}
i\frac{\partial }{\partial s} \left[\begin{array}{c}
\zeta_1(z) \\
\zeta_2^\ast(z)
\end{array}\right] = \left[\begin{array}{cc}
a_{11}(\Omega) & a_{12}(\Omega) \\
a_{21}(\Omega) & a_{22}(\Omega)
\end{array}\right]  \left[\begin{array}{c}
\zeta_1(z) \\
\zeta_2^\ast(z)
\end{array}\right],
\end{equation}
where $\zeta_j^\ast(z)$($j=1,2$) is the complex conjugate of the field $\zeta_j(z)$($j=1,2$), with the elements of the system's matrix $M$ being given by
\begin{equation}\label{eq11}
\begin{split}
a_{11} = & \frac{\beta_{2}}{2}\Omega^ 2 +  \frac{\beta_{4}}{24}\Omega^4 
+  \frac{\beta_{6}}{720} \Omega^6+\alpha P_0-\chi P_0\Omega^2,\\
a_{12} =& \alpha P_0-\chi P_0\Omega^2,\;\;
a_{21} = -\left(\alpha P_0-\chi P_0\Omega^2\right),\\
a_{22} = & - \left(\frac{\beta_{2}}{2}\Omega ^ 2 +  \frac{\beta_{4}}{24}\Omega ^4 
+  \frac{\beta_{6}}{720} \Omega ^6+\alpha P_0-\chi P_0\Omega^2\right).
\end{split}
\end{equation}
\begin{figure}[!]
\centering
\includegraphics[width=3.70in]{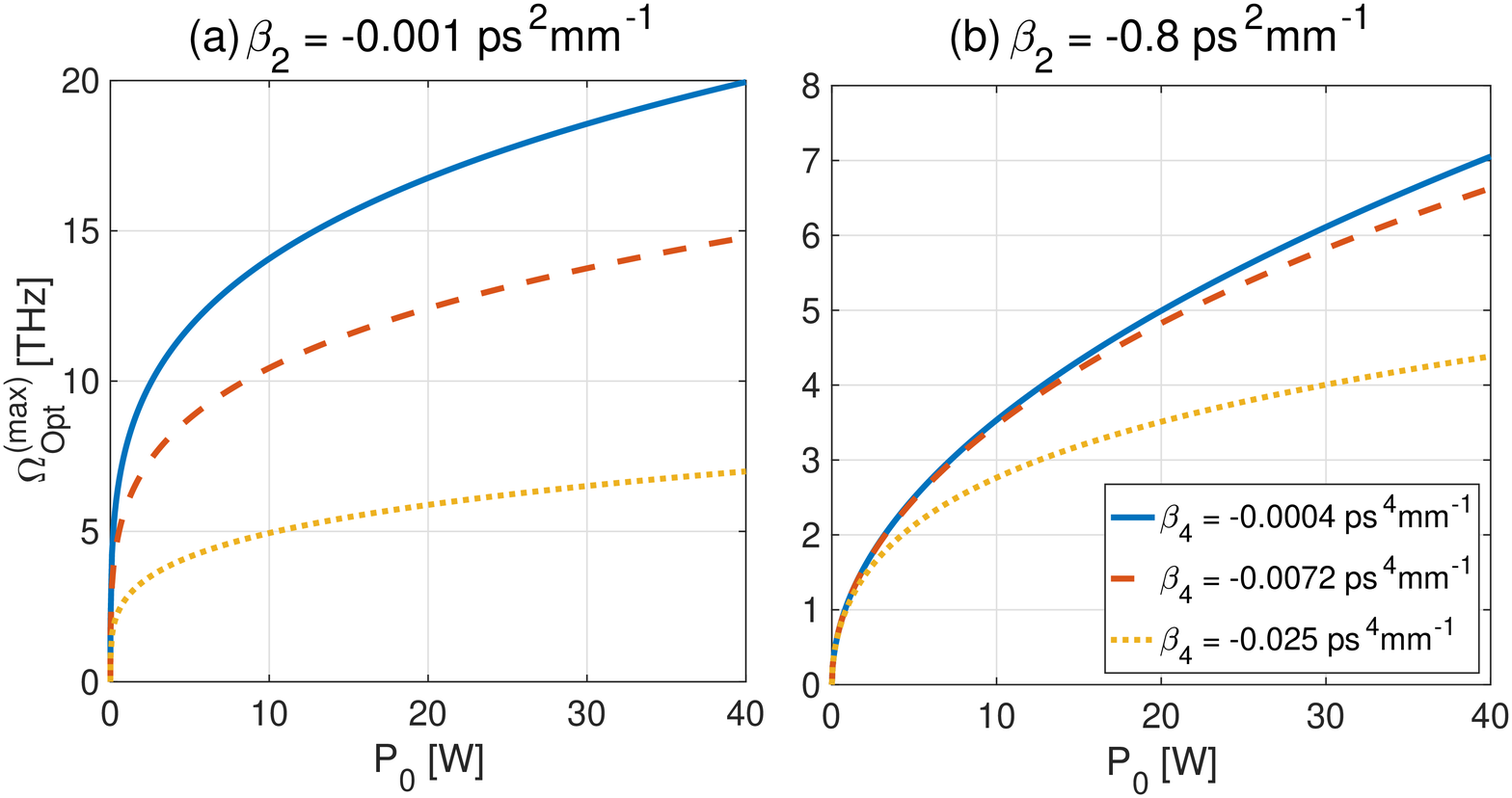}
\caption{Plots of the OMF versus the pump power $P_0$, computed from solution (\ref{eq15}). The panel (a) shows results for $\beta_2=-0.001$ps$^2$mm$^{-1}$ and panel (b) displays the OMF for $\beta_2=-0.8 $ps$^2$mm$^{-1}$, with $|\beta_4|$ taking increasing values in each of the cases.  The other parameter values are: $\alpha=0.5$W$^{-1}$.mm$^{-1}$  and $\beta_6=0$.}\label{fig1}
\end{figure}
The MI gain is investigated from the imaginary part of the wavenumber $k$ of the perturbation that can be obtained using the characteristic equation ${\rm Det}\left| M - k I\right|  = 0 $, with $I$ being a $2 \times 2$ identity matrix. One therefore obtains the dispersion relation
\begin{widetext} 
\begin{equation}\label{eq12}
\begin{split}
k =  \sqrt{\left(\alpha P_0+m\Omega ^ 2 +  \frac{\beta_{4}}{24}\Omega ^4 +  \frac{\beta_{6}}{720} \Omega ^6\right)^2-\left(\alpha P_0-\chi P_0\Omega^2\right)^2},
\end{split}
\end{equation}
where $m=({\beta_{2}}/{2}-\chi P_0)$. The MI will take place under the condition where the wavenumber $k$ has a nonzero imaginary part. The latter will be responsible for an amplitude exponential growth. The power gain, through which the MI is predicted is such that $ \Lambda(\Omega ) = 2{\rm Im}(k)$, which is specifically obtained in the form
\begin{equation} \label{eq13}
\Lambda(\Omega) = 2\sqrt{\left(\alpha P_0-\chi P_0\Omega^2\right)^2
-\left(\alpha P_0+m\Omega ^ 2 +  \frac{\beta_{4}}{24}\Omega ^4 +  \frac{\beta_{6}}{720} \Omega ^6\right)^2}.
\end{equation}
\end{widetext}
Interestingly, it is obvious that the above MI gain depends on all the involved terms, including the nonlocal one. To carry out our analysis, the MI gain will be parametrically studied using different combinations of parameters, especially the higher-order dispersions and the nonlocal contribution to the nonlinearity. In the meantime, the Optimum modulation frequency (OMF) can be obtained from the above expression of the MI gain by solving the equation $d\Lambda/d\Omega=0$. This leads to the equation:
\begin{equation}\label{eq14}
\begin{split}
&\frac{\beta _6^2 }{43200}\Omega ^{10}+\frac{ \beta _4 \beta _6}{864} \Omega ^8+\left(\frac{\beta _4^2}{72}+\frac{\beta _6 m}{45}\right)\Omega ^6 \\
&+\left(\frac{\beta _4 m}{2}+\frac{ \alpha  \beta _6 P_0}{60}\right)\Omega ^4-\left(4 P_0^2 \chi ^2-4 m^2-\frac{ \alpha  \beta _4 P_0}{3}\right)\Omega ^2\\
&+4 \alpha   P_0\left(m+P_0\chi \right)=0.
\end{split}
\end{equation}
Eq.(\ref{eq14}) constitutes a higher-order polynomial equation whose solutions cannot easily be found analytically. Even with softwares like  Mathematica or Matlab, only solutions related to specific choices of parameters can be obtained using symbolic computations. In what follows, we address some of such accessible cases that will allow us to qualitatively study the impact of the different constituents on the MI gain.
\begin{figure}[!]
\centering
\includegraphics[width=3.70in]{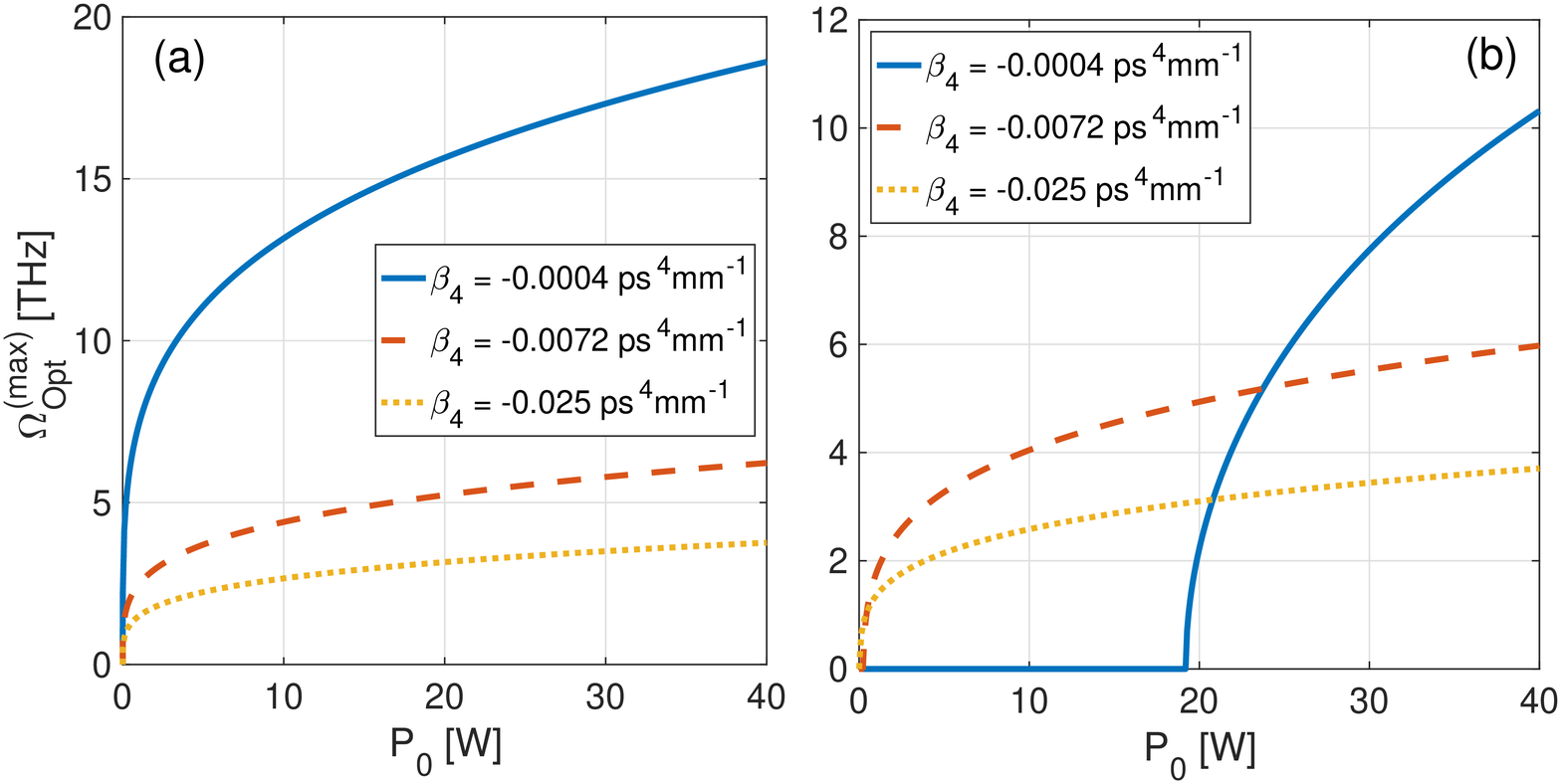}
\caption{Plots of the OMF as a function of the pump power $P_0$ for $\alpha=-0.5$W$^{-1}$.mm$^{-1}$, obtained from Eq.(\ref{eq15}). In panel (a), the second-order dispersion is switched off, i.e., $\beta_2=0$, while in panel (b) $\beta_2$ should take weak values for the OMF to exist. The values chosen in this case is $\beta_2=-0.8 $ps$^2$mm$^{-1}$, with $|\beta_4|$ taking increasing values in each of the cases, the rest of the parameters are such that $\gamma=0$, and $\beta_6=0$.}\label{fig2}
\end{figure}

\subsection{Optimum modulation frequency and MI gain spectrum}
\subsubsection{The optimum modulation frequency}
On solving Eq.(\ref{eq14}), it is possible to find frequency modes from which all the physical processes underlying the MI mechanism can be obtained. This will be possible only if the different frequencies are real and positive. One should also be aware that depending on the parameter choices, mainly the dispersion coefficients $\beta_2$, $\beta_4$, and $\beta_6$, one might get different results as is clearly addressed here, through the following cases.\\
\\
{\bf (i)} {\bf Case $\beta_6=0$, $\beta_4\neq0$, $\beta_2\neq0$, and $\chi=0$ }\\
\\
This particular case addresses the situation where the nonlocal term and the sextic dispersion term are switched off. Eq.(\ref{eq14}), therefore,  admits six solutions, with only the following being likely to be real and positive, depending on the parameter choices:
\begin{equation}\label{eq15}
\begin{split}
\Omega_{\rm Opt}=\sqrt{-\frac{12 m}{\beta _4}\pm\frac{2\sqrt{6} \sqrt{6 m^2-\alpha  \beta _4 P_0}}{\beta _4}}.
\end{split}
\end{equation}
The corresponding features are displayed in Fig.~\ref{fig1}, where the OMF is plotted versus the pump power $P_0$ with $|\beta_2|$ taking increasing values. In Fig.~\ref{fig1}(a), where we have fixed $|\beta_2|=0.001$ps$^2$mm$^{-1}$, the action of the quartic dispersion is perceptible, due to the weak value of the quadratic dispersion. Moreover, the OMF is a decreasing function of $|\beta_4|$, a behavior that is also perceptible in Fig.~\ref{fig1}(b), where $|\beta_2|=0.8$ps$^2$mm$^{-1}$. However, the magnitude of the OMF drastically drops, while it is not dramatically sensitive to the effect of increasing $|\beta_4|$. In the particular case where the quadratic dispersion and the nonlocality are completely switched off, the OMF reduces to $\Omega_{\rm Opt}=(-72\alpha P_0/\beta_4)^{1/4}$. Such a critical case, for $\alpha=-0.5$W$^{-1}$.mm$^{-1}$, is shown in Fig.~\ref{fig2}(a), where the OMF is extracted as a function of the pump power. When $|\beta_4|$ increases as previously, the OMF drops. Under the same conditions but with $|\beta_2|\neq 0$, its value should be strong enough for the function of the pump to be effective in MI onset. This is shown in Fig.~\ref{fig2}(b), where  $|\beta_4|=0.0004$ps$^4$mm$^{-1}$ excludes some interval of $P_0$ after which the OMF exists. For the remaining values of $\beta_4$, the OMF drops as previously, while the instability gets preserved.\\
\\
\begin{figure}[t]
\centering
\includegraphics[width=3.70in]{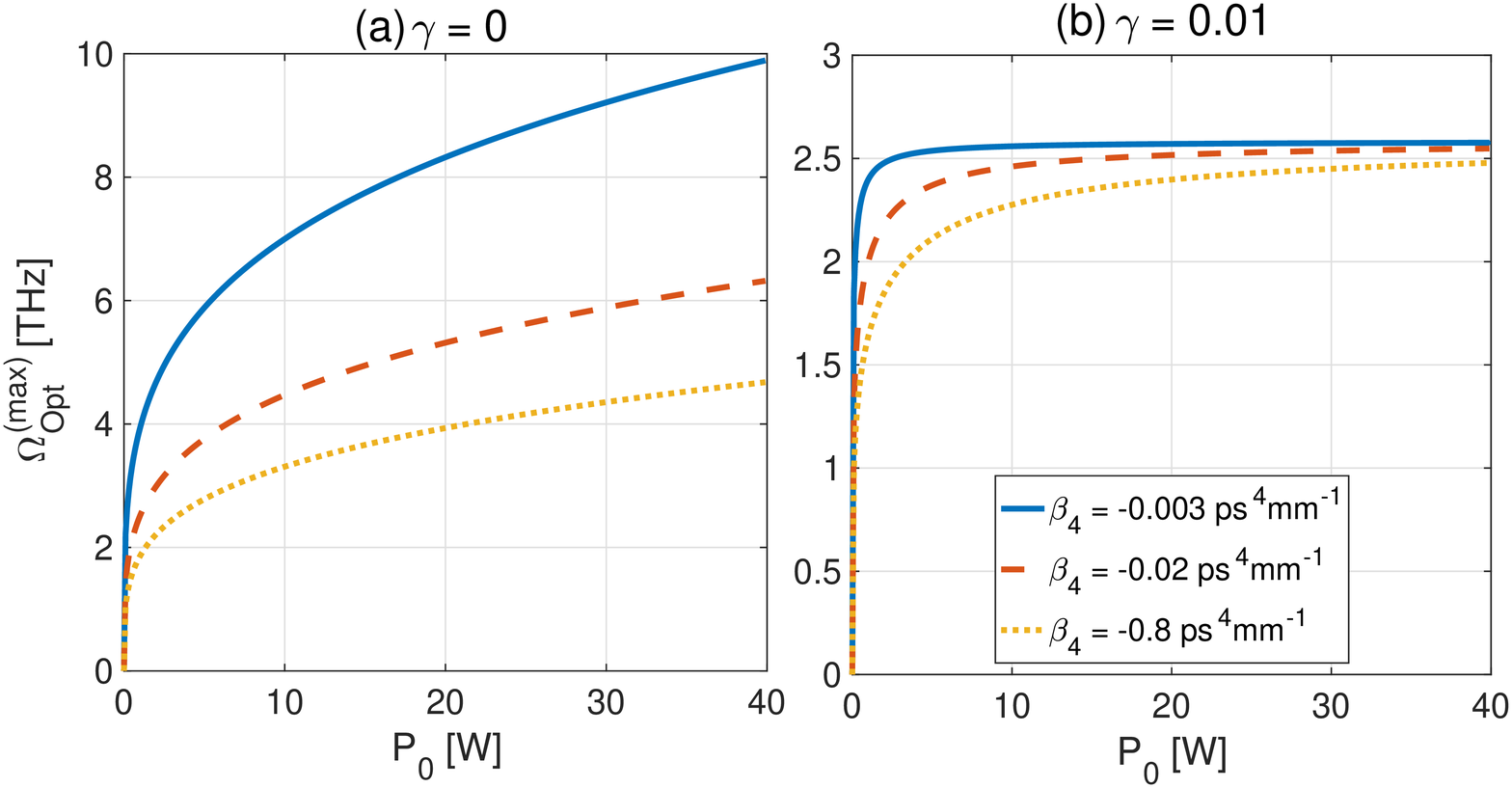}
\caption{Plots of the OMF versus the pump power $P_0$ in the absence [panel (a)] and presence [panel (b)] of the nonlocal term. Computed from the OMF formula (\ref{eq15}), each of the cases is studied for the fourth-order dispersion coefficient taking the respective values $\beta_4=-0.003$ps$^4$mm$^{-1}$ (solid blue line), $\beta_4=-0.02$ps$^4$mm$^{-1}$ (dashed red line) and $\beta_4=-0.8$ ps$^4$mm$^{-1}$ (dotted yellow line). The other parameter values are: $\alpha=0.5$W$^{-1}$.mm$^{-1}$, $\beta_6=0$, and $\beta_2=0$.}\label{fig3}
\end{figure}
\begin{figure}[t]
\centering
\includegraphics[width=3.70in]{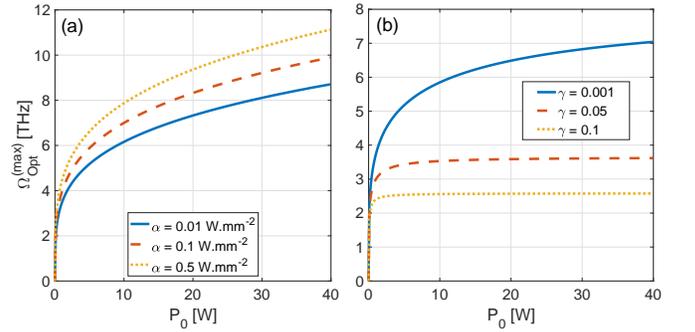}
\caption{The OMF of Eq.(\ref{eq15}) is plotted versus the pump power for (a): the nonlinearity coefficient changing and (b) the nonlocality parameter changing. In panel (a), the nonlocality parameter is taken to zero and $\alpha$ takes the respective values $\alpha=0.01$ W$^{-1}$.mm$^{-1}$ (solid blue line), $\alpha=0.1$ W.mm$^{-2}$ (dashed red line) and $\alpha=0.5$ W$^{-1}$.mm$^{-1}$ (dotted yellow line). To plot panel (b), we have fixed $\alpha=0.5$ W$^{-1}$.mm$^{-1}$ and the nonlocality parameter has been given the values $\gamma=0.001$ (solid blue line), $\gamma=0.05$ (dashed red line) and $\gamma=0.1$ (yellow dotted line). The other parameter values are:  $\beta_6=0$, $\beta_4=-0.8$ps$^4$mm$^{-1}$,  and $\beta_2=0$.}\label{fig4}
\end{figure}
\begin{figure}[t]
\centering
\includegraphics[width=3.70in]{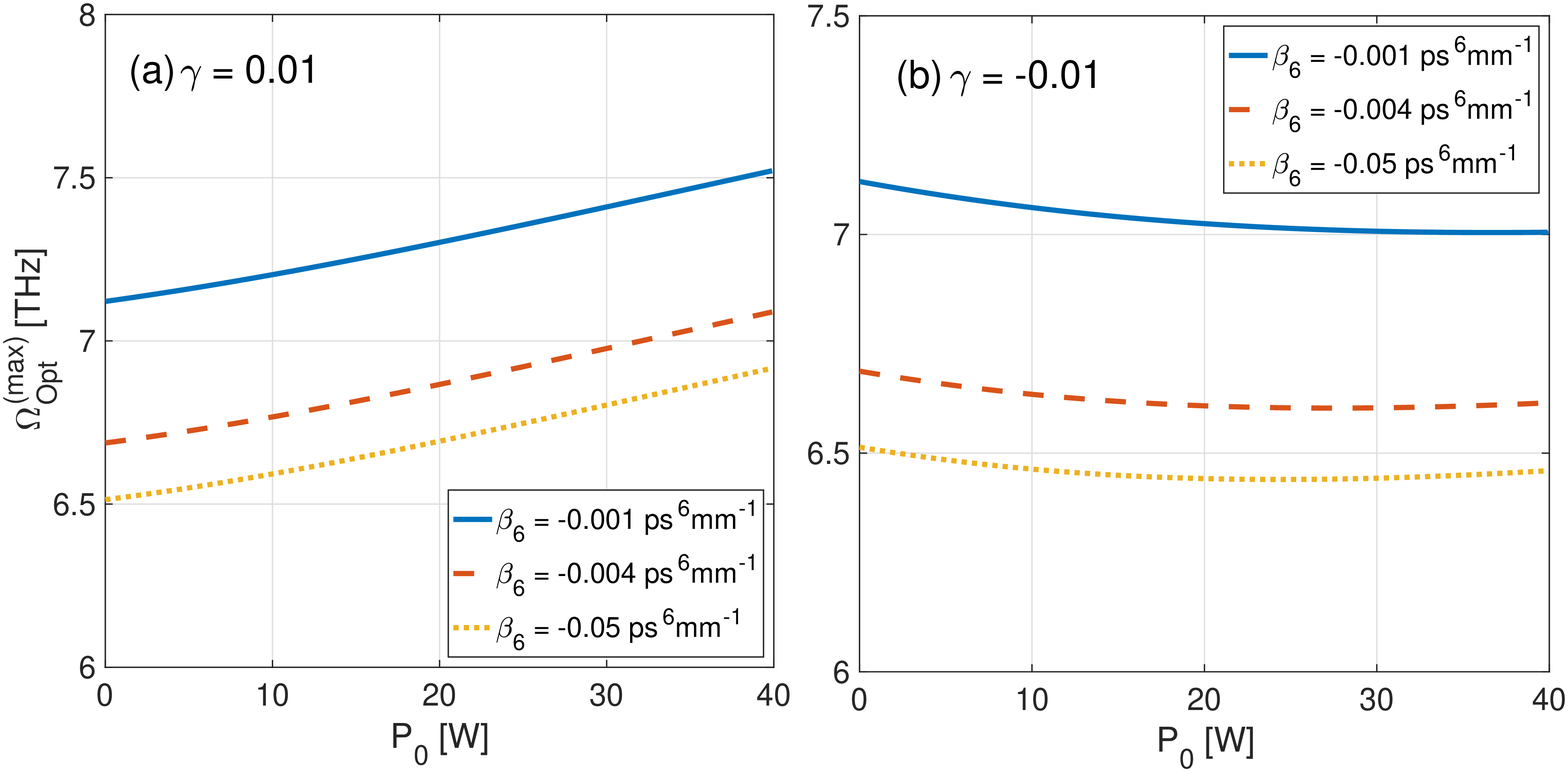}
\caption{The OMF is numerically extracted and plotted versus the pump power in the context where the $\beta_4=0$, with $\beta_2=-0.8$ps$^2$mm$^{-1}$ and the sextic dispersion taking the values $\beta_6=-0.001$ps$^6$mm$^{-1}$  (solid blue line), $\beta_6=-0.004$ps$^6$mm$^{-1}$ (dashed red line) and $\beta_6=-0.004$ps$^6$mm$^{-1}$ (dotted yellow line) . In panel(a), the nonlocality parameter is given the value $\gamma=0.01$, while in panel (b) it takes the value $\gamma=-0.01$, with $\alpha=0.1$ W$^{-1}$.mm$^{-1}$.}\label{fig5}
\end{figure}
\begin{figure}[t]
\centering
\includegraphics[width=3.70in]{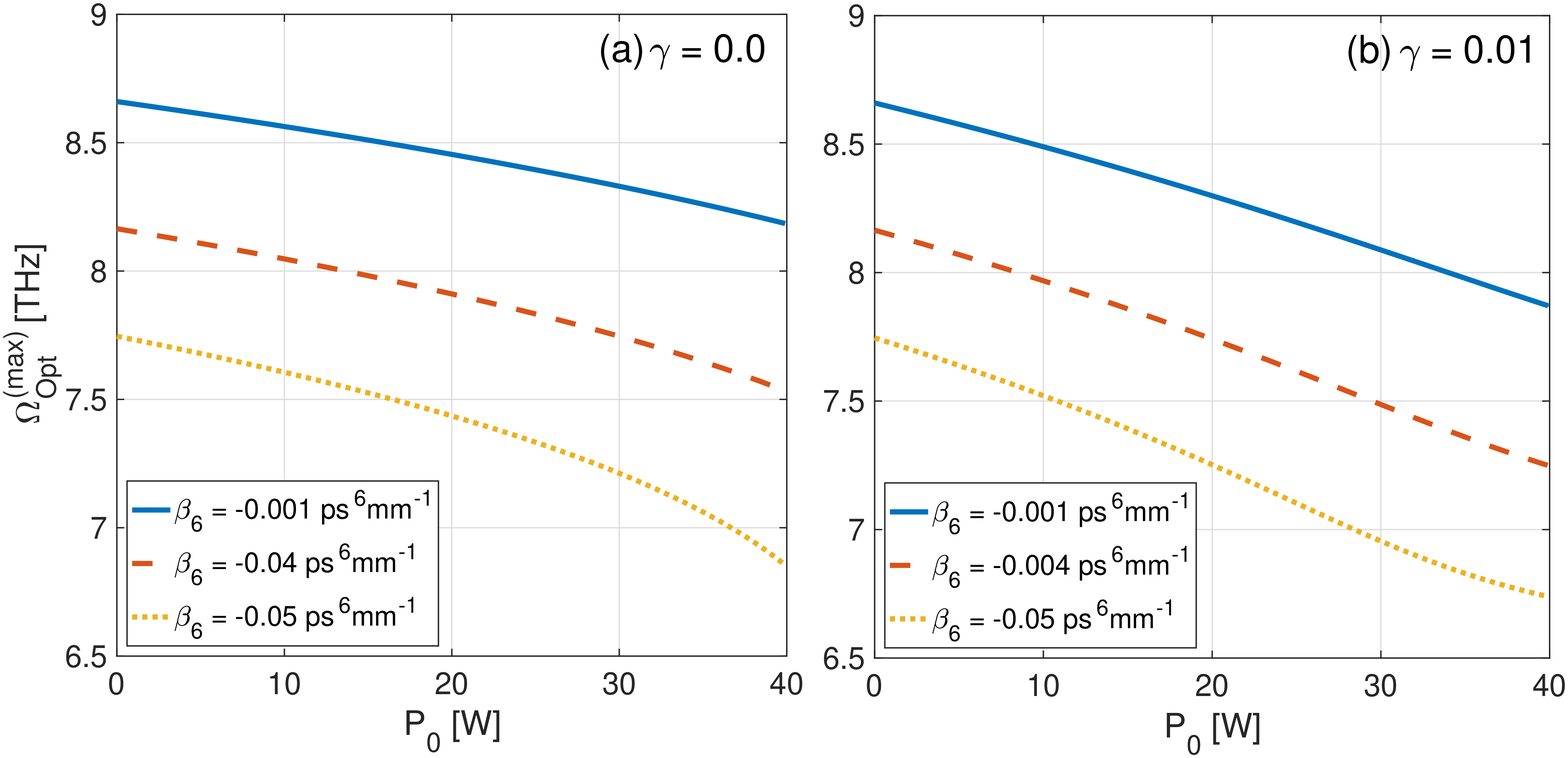}
\caption{The OMF  is numerically extracted and plotted versus the pump power for (a): the nonlinearity coefficient changing and (b) the nonlocality parameter changing. In panel (a), the nonlocality parameter is taken to zero and $\alpha$ takes the respective values $\alpha=0.01$ W.mm$^{-2}$ (solid blue line), $\alpha=0.1$ W$^{-1}$.mm$^{-1}$ (dashed red line) and $\alpha=0.5$ W$^{-1}$.mm$^{-1}$ (dotted yellow line). To plot panel (b), we have fixed $\alpha=0.5$ W$^{-1}$.mm$^{-1}$ and the nonlocality parameter has been given the values $\gamma=0.001$ (solid blue line), $\gamma=0.05$ (dashed red line) and $\gamma=0.1$ (yellow dotted line). The other parameter values are:  $\beta_6=0$, $\beta_4=-0.8$ps$^4$mm$^{-1}$,  and $\beta_2=0$.}\label{fig6}
\end{figure}
{\bf (ii)} {\bf Case $\beta_6=0$, $\beta_4\neq0$, $\beta_2=0$, and $\chi\neq0$}\\
\\
In this case, the sextic and quadratic dispersions are canceled, while the nonlocality is existent, leading to the solution
\begin{equation}\label{eq16}
\Omega_{\rm Opt}=\sqrt{\frac{18 P_0 \chi }{\beta _4}\pm\frac{2\sqrt{3} \sqrt{P_0 \left(27 P_0 \chi ^2-2 \alpha  \beta _4\right)}}{\beta _4}}
\end{equation}
for Eq.(\ref{eq14}). The results related to this case are delivered by the plots of Fig.~\ref{fig3}(a), where the nonlocality is switched off, with $\beta_2=0$, which shows the possibility of MI. In presence of the nonlocality, the decreasing values of $|\beta_4|$ show a saturation behavior where after a transient increase, the values of the pump power that support the OMF seem to be uniform [see Fig.~\ref{fig3}(a)]. On the other hand, increasing the Kerr nonlinear coefficient enhances the OMF [see Fig.~\ref{fig4}(a)], and a saturation effect takes place when the nonlocality parameter increases in a context where the OMF drops with increasing $\gamma$ [see Fig.~\ref{fig4}(b)].

From the above two cases, it is obvious that the FOD, second-order dispersion and the nonlocality parameter importantly affect the occurrence of MI. As mentioned earlier, one of the problems faced here is solving cases involving the SOD coefficient, which may be addressed numerically. In what follows, we discuss some of such few cases.\\
\\
{\bf (iii)} {\bf Case $\beta_6\neq0$, $\beta_4=0$, $\beta_2\neq0$, and $\gamma\neq0$ }\\
\\
This case engages the sextic dispersion, the second-order dispersion, the Kerr nonlinearity, and the weak nonlocality. The corresponding results are displayed in Fig.~\ref{fig5}, where panel (a) is dedicated to the positive nonlocality parameter $\gamma=0.01$ under the competition between negative second-order and sextic dispersion coefficients. In general, there is a high sensitivity of the OMF versus the magnitude of $|\beta_6|$. MI is characterized by a perpetual frequency shift with the value of the pump power increasing  [see Fig.~\ref{fig5}(a)]. Such a shift, when $\gamma<0$, is still perceptible, except that the OMF is a decreasing function of $|\beta_6|$. This clearly shows that the two cases of $\gamma$ (positive or negative) may support MI but with different solitonic features.\\
\\
{\bf (iv)} {\bf Case $\beta_6\neq0$, $\beta_4\neq 0$, $\beta_2=0$, and $\gamma\neq0$ }\\
\\
In the absence of nonlocality, panel (a)  of Fig.~\ref{fig6} testifies to the possibility of MI, especially when the quartic and sextic dispersions coexist in the absence of the second-order dispersion. The OMF shows high sensitivity against the magnitude of $|\beta_6|$, which, however, decreases with the pump power $P_0$ decreasing. The same scenario is ostensible in Fig.~\ref{fig6}(b), where the nonlocality is taken into consideration, with the other conditions remaining the same. This shows a competition between the dispersive nonlinearity and Kerr one and shows that when the sextic and quartic dispersions are well-chosen, the effect of the nonlocality is diluted by the presence of the Kerr nonlinearity, probably due to the weak factor $\chi=\alpha\gamma$ of the nonlocality.

Except for the few cases regarded above, other features may be explored using more combinations of the involved parameters. The study cannot, therefore, be exhaustive in that direction. However, the few studied cases can give the reader a flavor of what can be brought by including the nonlocality in such models, opening a new route for more comprehensive theoretical and experimental investigations. Nevertheless, to get more insight into the problem under study, the next step is devoted to studying the full MI gain using the above predictions, and we go beyond them to show the impact of combining system parameters on the MI gain. To that effect, the MI gain proposed in Eq.(\ref{eq13}) will be used, and calculations will be conducted in the case where $\beta_2$ is negative when considered.
\begin{figure}[t]
\centering
\includegraphics[width=3.70in]{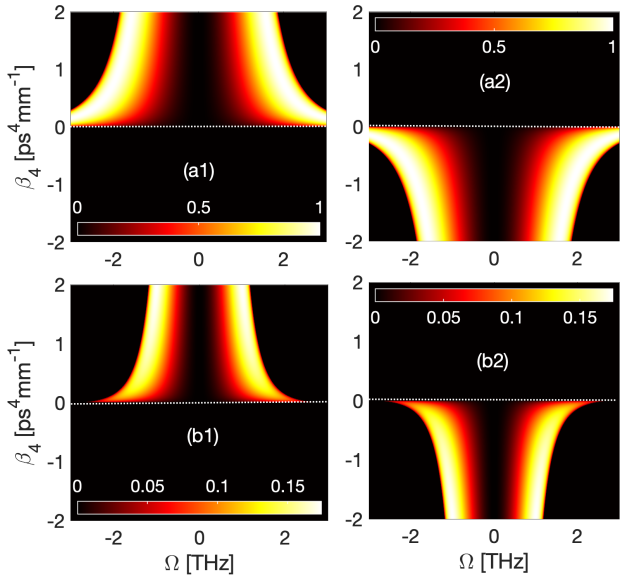}
\caption{The panels show the variations of the MI gain in the $(\Omega,\beta_4)-$plane in the absence of the second-order dispersion. Panels (aj)$_{j=1,2}$ are plotted in absence of nonlocality with $\alpha$ taking the values -0.5 W$^{-1}$.mm$^{-1}$ [panel (a1)] and 0.5 W.mm$^{-2}$ [panel (a2)]. In presence of nonlocality ($\gamma=0.01$), one obtains the features of panels (bj)$_{j=1,2}$, with the nonlinearity coefficient taking the respective values -0.5 W$^{-1}$.mm$^{-1}$ and 0.5 W$^{-1}$.mm$^{-1}$ for panels (b1) and (b2). The rest of the parameters are such that $\beta_6=0$, and $P_0=1$W.}\label{fig7}
\end{figure}
\begin{figure}[!]
\centering
\includegraphics[width=3.70in]{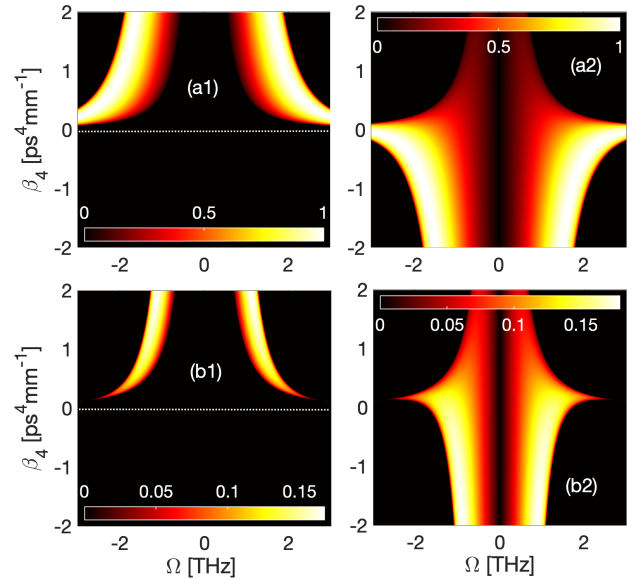}
\caption{The MI gain is represented in the $(\Omega,\beta_4)-$plane in presence of the in presence  of the second-order dispersion $\beta_2=-1.5$ps$^2$mm$^{-1}$. Panels (aj)$_{j=1,2}$ correspond to the case where the nonlocality is switched off, with the nonlinear coefficient taking the respective values $\alpha=-0.5$ W.mm$^{-2}$ and $\alpha=0.5$ W$^{-1}$.mm$^{-1}$ for panels (a1) and (a2). In panels (bj)$_{j=1,2}$, similar calculations are repeated but in presence of the nonlocal parameter that takes the value $\gamma=0.01$, with the other  parameters being $\beta_6=0$, and $P_0=1$ W.}\label{fig8}
\end{figure}
\begin{figure}[!]
\centering
\includegraphics[width=3.70in]{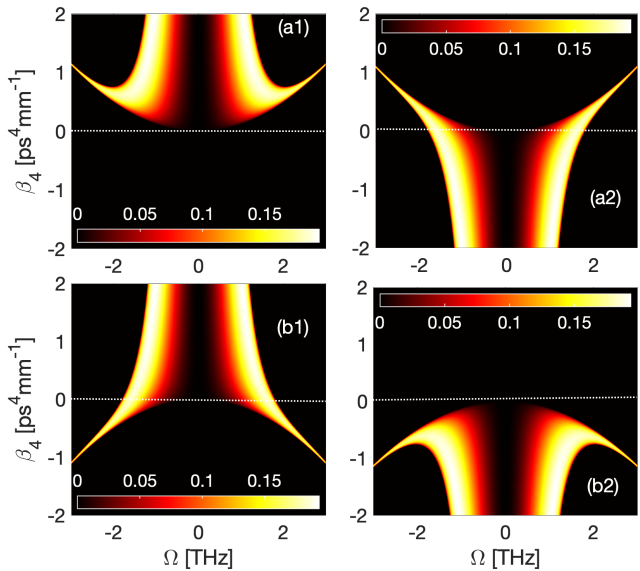}
\caption{The panels show the variations of the MI gain in the $(\Omega,\beta_4)-$plane in absence  of the second-order dispersion ($\beta_2=0$),  a non-nil nonlocality, $\gamma=0.01$. The effect of the sixth-order dispersion coefficient $\beta_6$ is confronted to the effect of the cubic nonlinearity for the emergence of the gain spectrum. Panels (aj) show $\Lambda(\Omega)$ for $\alpha=0.5$ W$^{-1}$.mm$^{-1}$, with $\beta_6=0.2$ ps$^6$mm$^{-1}$ [panel (a1)] and $\beta_6=-0.2$ ps$^6$mm$^{-1}$ [panel (a2)]. Panels (bj) display $\Lambda(\Omega)$ for $\alpha=-0.5$ W$^{-1}$.mm$^{-1}$, where $\beta_6=0.2$ ps$^6$mm$^{-1}$[panel (b1)] and $\beta_6=-0.2$ ps$^6$mm$^{-1}$ [panel (b2)], with the input power being $P_0=1$ W.}\label{fig9}
\end{figure}

\subsubsection{The MI gain spectrum}

In order to conduct exhaustively our MI analysis and predict regions of parameter for the occurrence of MI, the MI gain will be, in general, plotted versus the perturbation frequency $\Omega$ and the quartic dispersion coefficient $\beta_4$. 

The panels of Fig.~\ref{fig7} show the MI gain versus the perturbation frequency and the quartic dispersion coefficient $\beta_4$ for negative and positive cubic nonlinearity coefficient $\alpha$. This is carried out in the absence of second-order and sextic dispersions. In general,  the instability is manifested by two frequency sidebands that vanish at $\beta_4=0$. On the one hand, when $\alpha<0$, the instability takes part in the area $\beta_4>0$, while the MI emerges for $\beta_4<0$ when $\alpha>0$. We should stress that in this particular case, the nonlocality is switched off, and the windows of instability are symmetrically large, compared to the case where $\gamma\neq 0$, as shown in Fig.~\ref{fig7}(bj)$_{j=1,2}$, with $\gamma=0.01$. Obviously, the MI gain displays the same features, except that including the nonlocality shrinks the instability sidebands that, nevertheless, remain symmetric. Precision should, however, be made that the symmetric character of the MI gain is inherent to optical fibers relying on the NLS equation, whose MI gain is usually a quadratic function of the perturbation frequency $\Omega$. This has been reported even in the presence of other physical effects such as higher-order dispersion, higher-order nonlinearities, and cross-phase modulation~\cite{TabiOpLett2022}, to cite a few. In the meantime, recent works on materials have demonstrated that effects related to spin-orbit (SO) coupling, for example, can affect the symmetric features of the MI gain. This has been the case for Bose-Einstein condensates with helicoidal SO coupling in one- and two-space dimensions~\cite{li2019modulational,TabiPLA2022a}, as well as Bose-Bose mixtures with helicoidal SO coupling and, more recently, with Rashba and Dresselhaus SO couplings~\cite{TabiPRA2021,TabiPLA2022}. Interestingly, the three types of SO coupling commonly include coupling through the first-order derivatives with respect to the space coordinate of the  BEC components. Referring to the optical fiber,  the appearance of symmetric sidebands is also referred to as Stokes, and anti-Stokes sidebands, which for some input powers, can become asymmetric~\cite{DindaOpCom2006}. For now, the nonlocality contributes to reducing the frequency band gap, while the Stokes and anti-Stokes sidebands get enhanced.  An important remark to be made here is that under the conditions where the nonlocality is off, the system, whose MI gain is shown in Fig.~\ref{fig7}, is nonintegrable. However, the stability and existence of some pure-quartic solitons with temporal profiles characterized by exponentially decaying tails, with additional oscillations, were comprehensively discussed in Ref.~\cite{TamPRA2020} for $\beta_4<0$. Nevertheless, a conclusion was reached that such entities were not really solitons and needed to be properly characterized. 

Beyond the above-discussed scenarios, one of the systems that were recently given attention is the modified NLS equation in the presence of quadratic and quartic dispersions. In presence of the nonlocality contribution, the figures of instability are displayed in Fig.~\ref{fig8}, where panels (aj)$_{j=1,2}$ show $\Lambda(\Omega)$ for  $\gamma=0$ and $\beta_2=-1.5$ ps$^2$mm$^{-1}$. For $\alpha=-0.5$ W$^{-1}$.mm$^{-1}$ [panel(a1)], the instability is supported by two symmetrical lobes showing the MI gain exclusively in the area $\beta_4>0$, while the maximum MI gain takes place in the area $\beta_4<0$ for $\alpha=0.5$ W$^{-1}$.mm$^{-1}$ and decreases when $\beta_4$ increases from 0 to infinity. In presence of the nonlocality, i.e., $\gamma=0.01$, the features delivered by Fig.~\ref{fig8}(bj)$_{j=1,2}$ are similar to those of panels (aj)$_{j=1,2}$, except that the sidebands get shrunk, and the instability tends to be restrained to the allocated frequency bands. Also, for $\alpha<0$, the band gap gets extended while it gets reduced under negative nonlinear effects. The solution of such a system is well-known, especially when the system is clear of the nonlocality effect, as initially proposed by Karlsson and H\"o\"ok~\cite{KarlssonOptCom1994} (when $\beta_2<0$ and $\beta_4<0$) and used by several authors under different contexts.
\begin{figure}[t]
\centering
\includegraphics[width=3.70in]{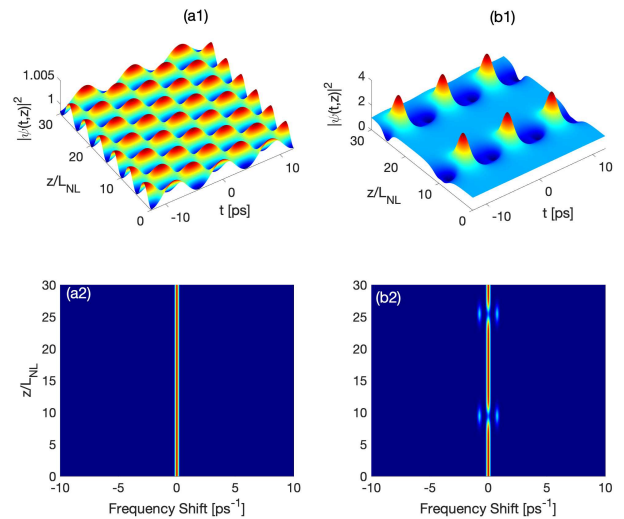}
\caption{Propagation of the perturbed plane wave intensity and wave pattern formation in presence of only the quadratic dispersion. Panels (aj)$_{j=1,2}$ correspond to the case where the nonlocality is switched off ($\gamma=0$), while panels (bj)$_{j=1,2}$ are obtained in presence of the nonlocality ($\gamma=0.01$). The rest of the parameters are such that: $\beta_6=0$, $\beta_4=-0.24$ ps$^4$mm$^{-1}$ , $\beta_2=0$, $\beta_6=0$, $\Omega_m=0.75$ THz, $P_0=1$W, and $\alpha=0.5$ W$^{-1}$.mm$^{-1}$. The upper line shows the evolution of the perturbed plane wave, while the lower line displays the spectral amplitude versus the propagation distance $z$ and the frequency shift.}\label{fig10}
\end{figure}
\begin{figure}[t]
\centering
\includegraphics[width=3.70in]{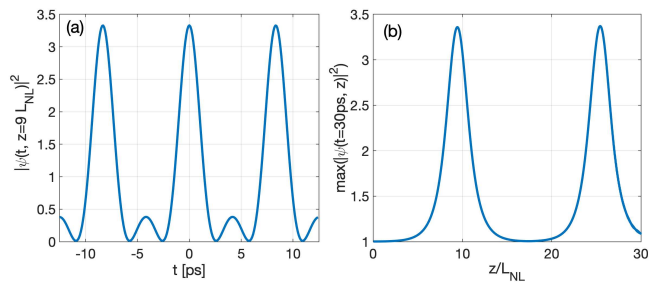}
\caption{Panel (a) shows a section of Fig.~\ref{fig10}(a) confirming the disintegration of the plane wave solution into trains of solitonic objects while panel (a) displays the maximum of the wave amplitude versus the propagation distance $z$. This happens in presence of the nonlocality which shows to support instability in pure-quartic NLS equation with parameters values being: $\gamma=0.01$,  $\beta_6=0$, $\beta_4=-0.2$ ps$^4$mm$^{-1}$, $\beta_2=0$, $\Omega_m=0.75$ THz, $P_0=1$W, and $\alpha=0.5$ W$^{-1}$.mm$^{-1}$.}\label{fig11}
\end{figure}
 \begin{figure}[t]
\centering
\includegraphics[width=3.70in]{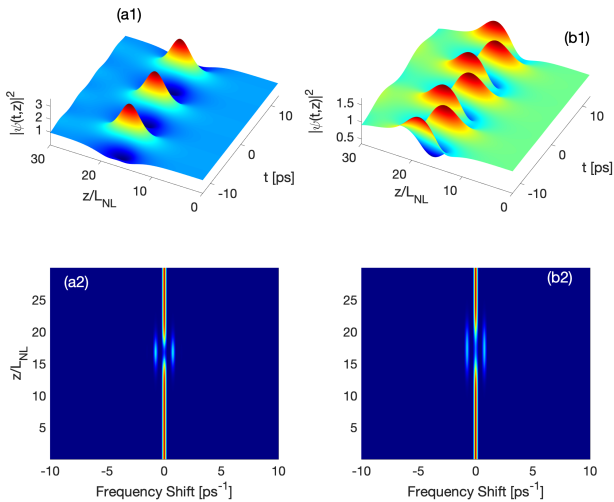}
\caption{Propagation of the perturbed plane wave intensity and wave pattern formation in absence of the nonlocality [panels (aj)$_{j=1,2}$] and in presence of the nonlocality [panel (bj)$_{j=1,2}$]. Panels (b1) and (b2) show the disintegration of the CW in the frequency-space for $\gamma=0$ and $\gamma=0.01$, respectively, with $\beta_6=0$, $\beta_4=-0.2$ ps$^4$mm$^{-1}$, $\beta_2=-1.5$ ps$^2$mm$^{-1}$, $\Omega=0.75$ THz, $P_0=1$W, and $\alpha=0.5$ W$^{-1}$.mm$^{-1}$. The upper line shows the evolution of the perturbed plane wave, while the lower line displays the spectral amplitude versus the propagation distance $z$ and the frequency shift.}\label{fig12}
\end{figure}

A special case of the MI gain is displayed in Fig.~\ref{fig9}, where the sextic dispersion, materialized by $\beta_6$, is confronted with the other dispersion orders, in presence of the nonlocality. We should stress that recent investigations have addressed the type of bright soliton that may exist in such a system, free from the nonlocality effect~\cite{QiangPRA2022}. In the meantime, the features of Fig.~\ref{fig9} reveal the existence of MI in such a model up to the extent the nonlocality is accommodated by the process of wave pattern generation. In this particular case, we have switched off the effect of the quadratic dispersion $\beta_2$, and panels (aj)$_{j=1,2}$ have been obtained for $\alpha=0.5$ W$^{-1}$.mm$^{-1}$, with $\beta_6$ taking the respective values $0.2$ ps$^6$mm$^{-1}$ and $-0.2$ ps$^6$mm$^{-1}$. Here, the MI gain displays a set of two symmetrical sidebands, but the effect of the negative sextic dispersion brings about re-entrant instability tails, with the whole gain pattern being restrained to the $\beta_4>0$ area [see Fig.~\ref{fig9}(a1)]. In Fig.~\ref{fig9}(a2), the previous conditions are kept, except that $\beta_6=-0.2$ ps$^6$mm$^{-1}$ shifts the major lobes of instability to the area $\beta_4<0$ along with extended symmetrical tails of instability in the area $\beta_4>0$. A reversed instability scenario appears for $\alpha=-0.5$W$^{-1}$.mm$^{-1}$, where curved instability tails are present for $\beta_6=-0.2$ ps$^6$mm$^{-1}$ [see Fig.~\ref{fig9}(b2)] and  extended tails for $\beta_6=0.2$ ps$^6$mm$^{-1}$ [see Fig.~\ref{fig9}(b1)]. 
\begin{figure}[!]
\centering
\includegraphics[width=3.70in]{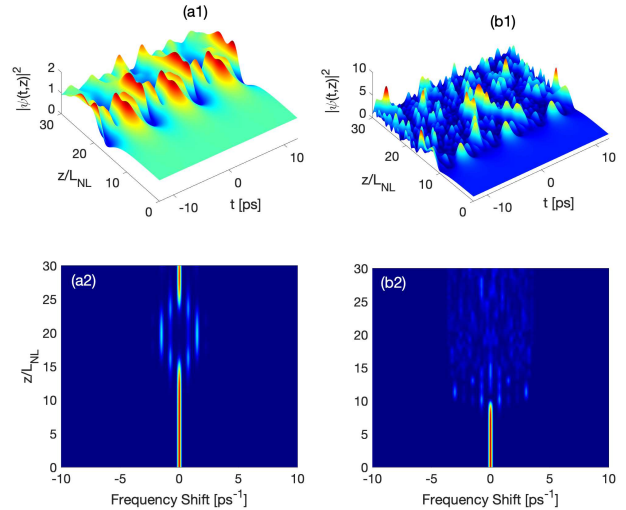}
\caption{Propagation of the perturbed plane wave intensity and wave pattern formation in presence of nonlocality for (aj)$_{j=1,2}$: $\beta_4=-0.01$ ps$^4$mm$^{-1}$, $\beta_2=-0.8$ ps$^2$mm$^{-1}$ and (bj)$_{j=1,2}$:  $\beta_4=-0.2$ ps$^4$mm$^{-1}$, $\beta_2=-0.8$ ps$^2$mm$^{-1}$, with $\gamma=0.01$, $\beta_6=0$, $\Omega=0.75$ THz, $P_0=1$W, and $\alpha=0.5$ W$^{-1}$.mm$^{-1}$. The upper line shows the evolution of the perturbed plane wave, while the lower line displays the spectral amplitude versus the propagation distance $z$ and the frequency shift.}\label{fig13}
\end{figure}
\begin{figure}[t]
\centering
\includegraphics[width=3.70in]{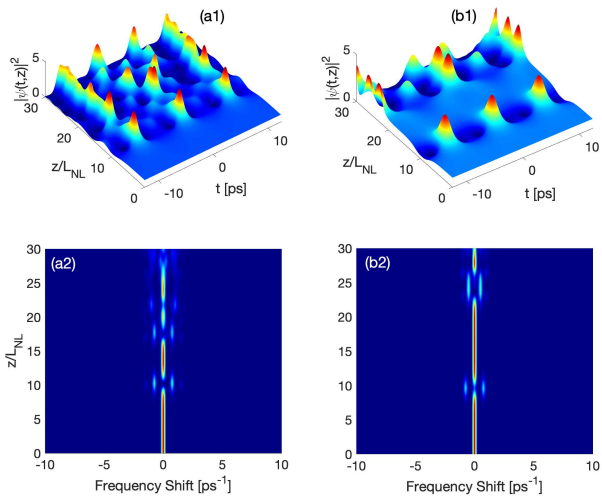}
\caption{Propagation of the perturbed plane wave intensity and wave pattern formation in absence of nonlocality for (aj)$_{j=1,2}$: $\beta_4=-0.01$ ps$^4$mm$^{-1}$, $\beta_2=-0.8$ ps$^2$mm$^{-1}$ and (bj)$_{j=1,2}$:  $\beta_4=-0.2$ ps$^4$mm$^{-1}$, $\beta_2=-0.8$ ps$^2$mm$^{-1}$, with $\gamma=0.01$, $\beta_6=0$, $\Omega=0.75$ THz, $P_0=1$W, and $\alpha=0.5$ W$^{-1}$.mm$^{-1}$. The upper line shows the evolution of the perturbed plane wave, while the lower line displays the spectral amplitude versus the propagation distance $z$ and the frequency shift.}\label{fig14}
\end{figure}

\section{Numerical experiment}\label{s4}

The linear stability analysis has given us information about the onset of the instability using various combinations of system parameters. Although the number of parameters is not high compared to some recent NLS models, the information gathered from the previous section are a clear evidence of the possibility of MI in the studied model. However, the main drawback of the linear stability analysis is its limitation in giving reliable information about the long-distance propagation of the input CW signal, the reason being that it results from the linearization around the unperturbed CW. Therefore, to confirm the accuracy of our analytical predictions, Eq.(\ref{eq5}) has been integrated using the split-step Fourier method. An initial signal input of the form 
\begin{equation}\label{eq17}
\begin{split}
\psi(t,z=0)=\sqrt{P_0}+\varepsilon\cos(\Omega_m t),
\end{split}
\end{equation} 
has been injected, with $\Omega_m=0.75$ THz being the modulation frequency, $\varepsilon=10^{-3}$ and $P_0=1$ W. Although the quadratic dispersion coefficient can take both positive and negative values, the whole study will be conducted using $\beta_2<0$, while the other parameters will be varied according to predictions from the linear stability analysis. The results will be generally extracted in terms of the signal intensity  $I=|\psi(t,z)|^2$. Additionally, simulations will be performed using a propagation distance $z=30 L_{NL}$, where $L_{NL}$ represents a nonlinear length defined as $L_{NL}=(\alpha P_0)^{-1}$, with $P_0$ being the input power defined above.

\subsection{MI under pure quartic dispersion: $\beta_2=0$, $\beta_4\neq 0$ and $\beta_6=0$}
\begin{figure}[!]
\centering
\includegraphics[width=3.70in]{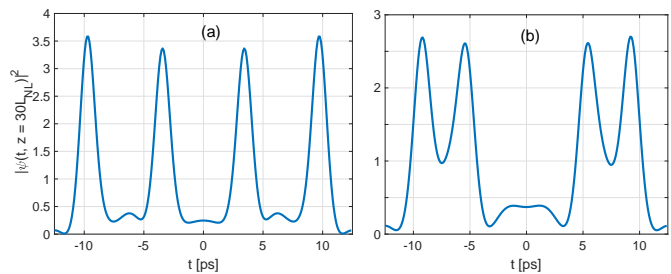}
\caption{Propagation of the perturbed plane wave intensity and wave pattern formation in absence of nonlocality for (a): $\beta_4=-0.01$ ps$^4$mm$^{-1}$, $\beta_2=-0.8$ ps$^2$mm$^{-1}$ and (b):  $\beta_4=-0.2$ ps$^4$mm$^{-1}$, $\beta_2=-0.8$ ps$^2$mm$^{-1}$, with $\gamma=0.01$, $\beta_6=0$, $\Omega_m=0.75$ THz, $P_0=1$W, and $\alpha=0.5$ W$^{-1}$.mm$^{-1}$.}\label{fig15}
\end{figure}

Considerable efforts have been made both analytically and numerically to understand the dynamics of pure-quartic solitons. In this context, from a theoretical point of view, the evolution of pure-quartic soliton has been studied using a Gaussian-like shape ansatz based on a variational approach. In the meantime, from the numerical point of view, it has been demonstrated that the dynamics of the pure-quartic soliton behaves like a temporal profile characterized by exponentially decaying tails with additional oscillations.  Only recently, close forms of pure-quartic solitons have been proposed in the forms of $sech$ and $tanh$ functions, which do not possess oscillating tails, markedly different from the conventional pure-quartic solitons in the absence of additional effects such as weak nonlocality~\cite{TrikiPLA2023}. This is also true in the present work when the nonlocality is not considered, as shown in Fig.~\ref{fig10}(aj)$_{j=1,2}$, where the CW under modulation remains stable, with  $\beta_2=-1.5$ ps$^2$mm$^{-1}$. In the presence of nonlocality, i.e., $\gamma\neq 0$, the instability develops, which shows that the non-integrability of the purely quartic NLS equation can be violated under additional input from other nonlinear terms, as depicted in Fig.~\ref{fig10}(bj)$_{j=1,2}$. Obviously, the MI is manifested by solitonic objects that form trains and have pulse-like shapes [see Fig.~\ref{fig11}]. Indeed, this confirms our analytical predictions and gives more credit to MI as a universal mechanism capable of leading to soliton formation in nonlinear systems when the balance between nonlinearity and dispersion is appropriate. Particularly,  the behaviors of Fig.~\ref{fig10}(b2) show some frequency shift, which confirms the emergence of sidebands. We should, however, stress that even though no exact solution can be found under the pure quartic dispersion, it was experimentally revealed by  Blanco-Redondo et al.~\cite{RedondoNatCom2016} that it may contribute to preserving the shape of any pulsating signal. Moreover, they proposed an approximate shape of the fundamental pure-quartic solution in the form of Gaussian and found agreement with their experimental observations. Other contributions have revealed that coupling negative quadric and quartic dispersions gives a possibility to solve the NLS equations, with a solution that was proposed by Karlsson and  H\"o\"ok~\cite{KarlssonOptCom1994}. The MI in such a situation is explored next.

\subsection{MI under quadratic and quartic dispersions: $\beta_2\neq 0$, $\beta_4\neq 0$ and $\beta_6=0$}

The results related to this case are shown in Figs.~\ref{fig12} and \ref{fig13}. The first case that is addressed here compares the situations where the nonlocality is absent and when it is taken into account. In the absence of the nonlocality, the competition between nonlinear and the available dispersion effects gives rise to the patterns of Fig.~\ref{fig12}(aj)$_{j=1,2}$. The instability takes place around the distance $z=15L_{NL}$ and manifests itself via the emergence of RW-like patterns. When the nonlocality is involved, the initial CW breaks into the patterns of Fig.~\ref{fig12}(bj)$_{j=1,2}$, where sneaky breathing patterns appear, with a reduced wave intensity. However, the frequency shift takes place over a longer distance compared to the events of Fig.~\ref{fig12}(aj)$_{j=1,2}$. This equally confirms our analytical predictions from which, when parameters are well-chosen inside the instability zone, the perturbation will grow exponentially and give rise to wave patterns.  Of course, we have restricted ourselves to the case of negative quadratic dispersion. The reader should, however, be reminded that our analytical calculations also predicted the existence of MI in the case where $\beta_2$ is positive. In that direction, Tam and Alexander~\cite{TamPRA2020} recently proposed a generalized solution that they named {\it meta-soliton} that was found for $\beta_2>0$ and $\beta_4<0$, with specific characteristics mostly applicable to laser systems. Moreover, single and multi-pulse solitary wave solutions to a generalized NLS equation, in presence of both quadratic and quartic dispersion terms were addressed by Parker and Aceves~\cite{ParkerPhysD2021}. The criteria for their existence and stability were also studied and it was revealed that multi-soliton structures were fundamentally unstable. In our context, another scenario relying on coupled quartic and quadratic dispersions is shown in Fig.~\ref{fig13}, where  $\beta_2$ takes the value $\beta_2=-0.8$ ps$^2$mm$^{-1}$, with $|\beta_4|$ increasing. In Fig.~\ref{fig13}(aj)$_{j=1,2}$, where we have fixed $\beta_4=-0.01$ ps$^4$mm$^{-1}$, the instability develops through the disintegration of the CW under a broad frequency shift as testified by Fig.~\ref{fig13}(a2). However, when $|\beta_4|$ is increased to 0.2 ps$^4$mm$^{-1}$, the CW disintegrates through a broad range of frequency sidebands, which identifies erratic dynamics of the instability. This is supported by highly localized structures as per the features of Fig.~\ref{fig13}(b2).  The patterns are also found to be very sensitive to the inclusion of the nonlocality, which further affects their dynamics when increased. For the patterns of Fig.~\ref{fig14}, the nonlocality parameter has been increased to $\gamma=0.08$, under the conditions where $\beta_2$ remains fixed and $|\beta_4|$ takes values  $0.01$ ps$^4$mm$^{-1}$ [see Fig.~\ref{fig14}(aj)$_{j=1,2}$] and $0.2$ ps$^4$mm$^{-1}$ [see Fig.~\ref{fig14}(bj)$_{j=1,2}$]. In the first case, after initiation, many frequency shifts take place as the patterns propagate, showing the long-distance robustness of MI [see Fig. ~\ref{fig14}(a2)]. On the contrary, for $\beta_4=-0.2$ ps$^4$mm$^{-1}$, the appearance of frequency shifts is reduced, although the initial step of MI is supported by the emergence of RWs. Moreover, the final stage of MI here is pictured in Fig.~\ref{fig15}, where panels (a), related to Fig~\ref{fig14}(aj)$_{j=1,2}$, shows a train of pulse and panel (b), extracted from Fig~\ref{fig14}(bj)$_{j=1,2}$, shows the disintegration of the CW into two-humped solitons.
\begin{figure}[t]
\centering
\includegraphics[width=3.50in]{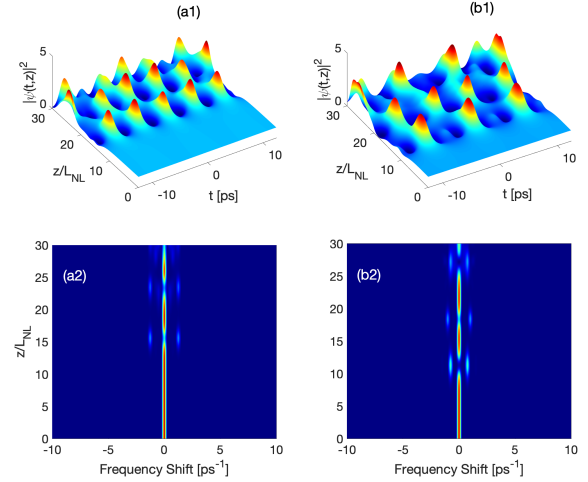}
\caption{Propagation of the perturbed plane wave intensity and wave pattern formation in absence of nonlocality for (aj)$_{j=1,2}$: $\beta_6=-0.1$ ps$^6$mm$^{-1}$, $\beta_4=-0.8$ ps$^4$mm$^{-1}$, $\beta_2=-1.5$ ps$^2$mm$^{-1}$ and (bj)$_{j=1,2}$:  $\beta_6=-0.5$ ps$^6$mm$^{-1}$, $\beta_4=-0.8$ ps$^4$mm$^{-1}$, $\beta_2=-1.5$ ps$^2$mm$^{-1}$, with $\gamma=0$, $\Omega_m=0.75$ THz, $P_0=1$W, and $\alpha=0.5$ W$^{-1}$.mm$^{-1}$. The upper line shows the evolution of the perturbed plane wave, while the lower line displays the spectral amplitude versus the propagation distance $z$ and the frequency shift.}\label{fig16}
\end{figure}
\begin{figure}[h]
\centering
\includegraphics[width=3.70in]{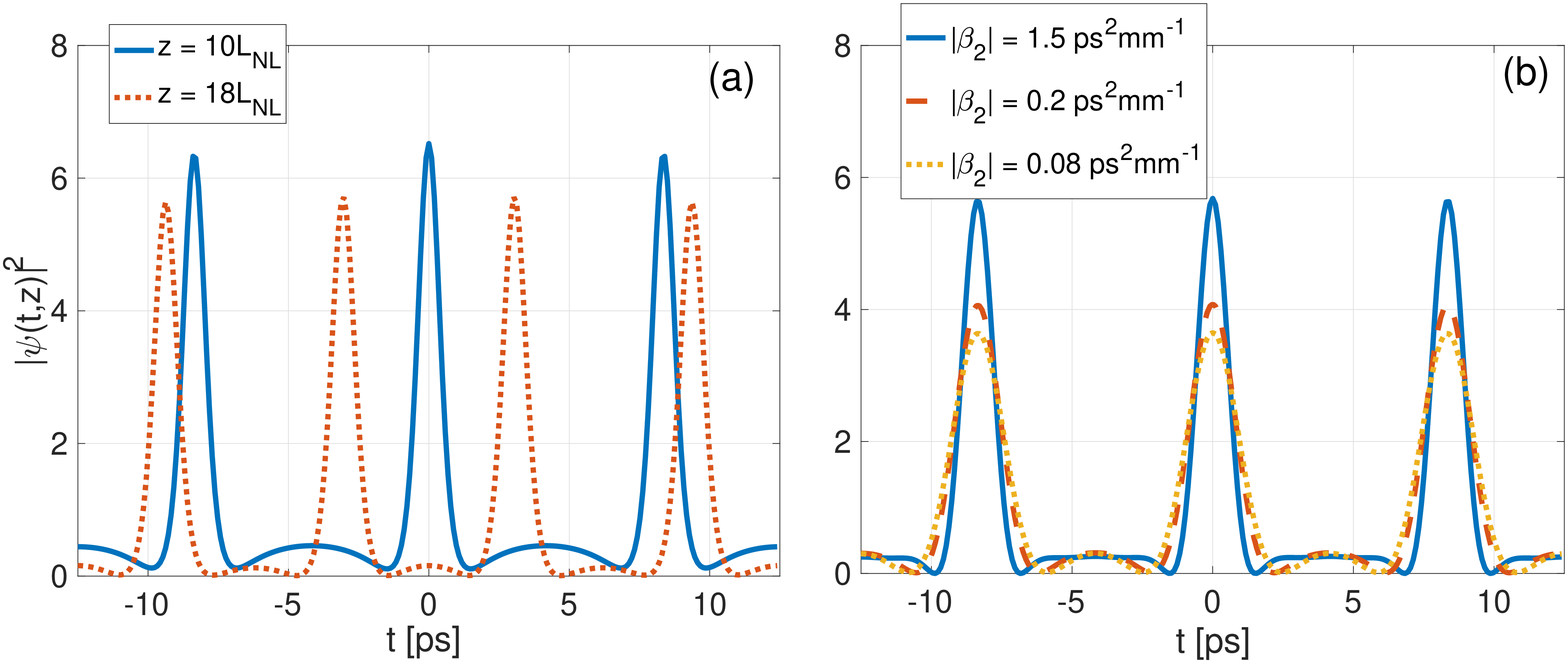}
\caption{Propagation of the perturbed plane wave intensity and wave pattern formation in absence of nonlocality for (a): $\beta_6=-0.1$ ps$^6$mm$^{-1}$, $\beta_4=-0.8$ ps$^4$mm$^{-1}$, $\beta_2=-1.5$ ps$^2$mm$^{-1}$ and (b):  $\beta_6=-0.5$ ps$^6$mm$^{-1}$, $\beta_4=-0.8$ ps$^4$mm$^{-1}$, $\beta_2=-1.5$ ps$^2$mm$^{-1}$, with $\gamma=0$, $\Omega=0.75$ THz, $P_0=1$W, and $\alpha=0.5$ W$^{-1}$.mm$^{-1}$.}\label{fig17}
\end{figure}

\subsection{MI under quadratic, quartic and sextic dispersions: $\beta_2\neq 0$, $\beta_4\neq 0$ and $\beta_6\neq 0$}

\begin{figure}[t]
\centering
\includegraphics[width=3.70in]{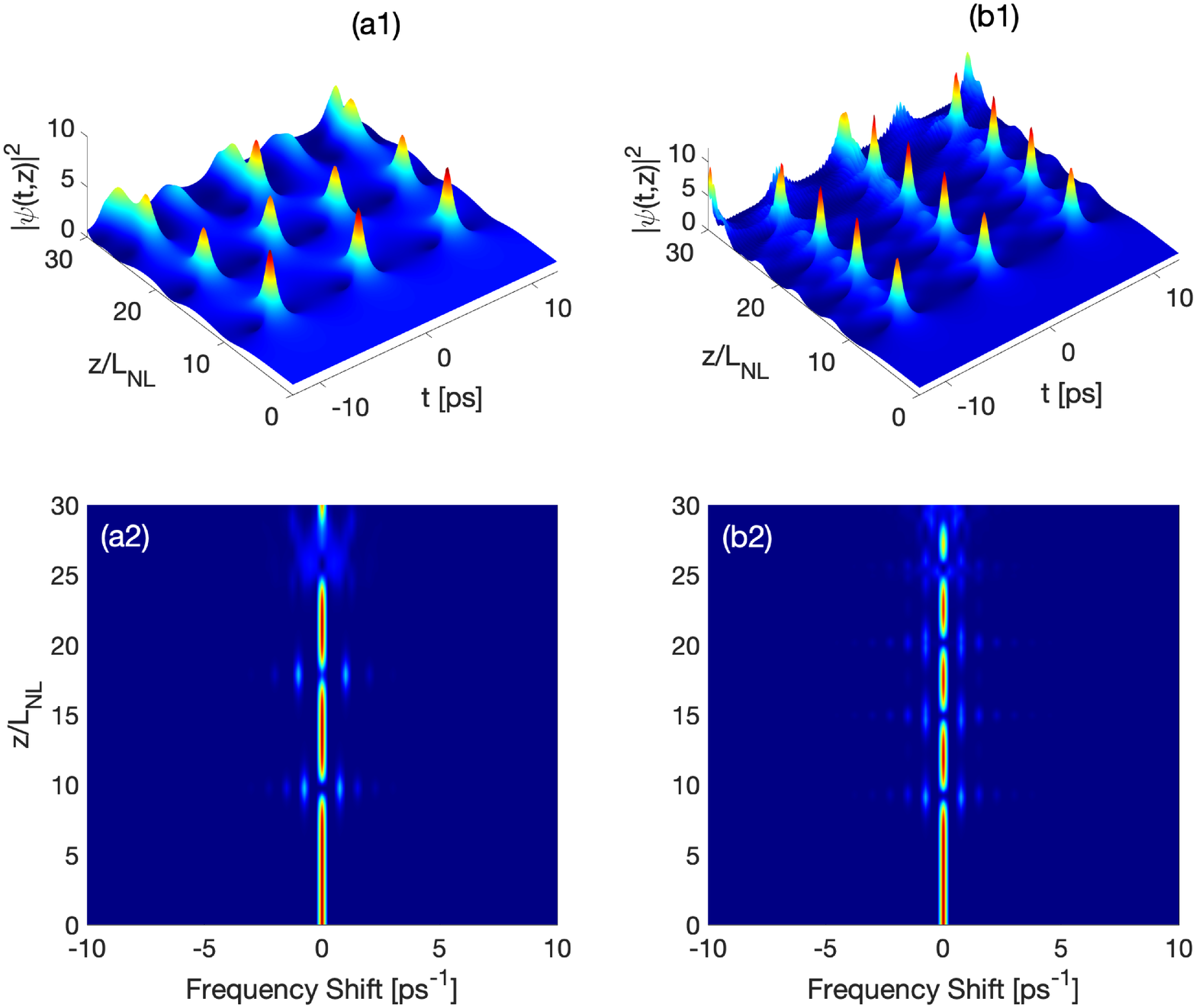}
\includegraphics[width=3.00in]{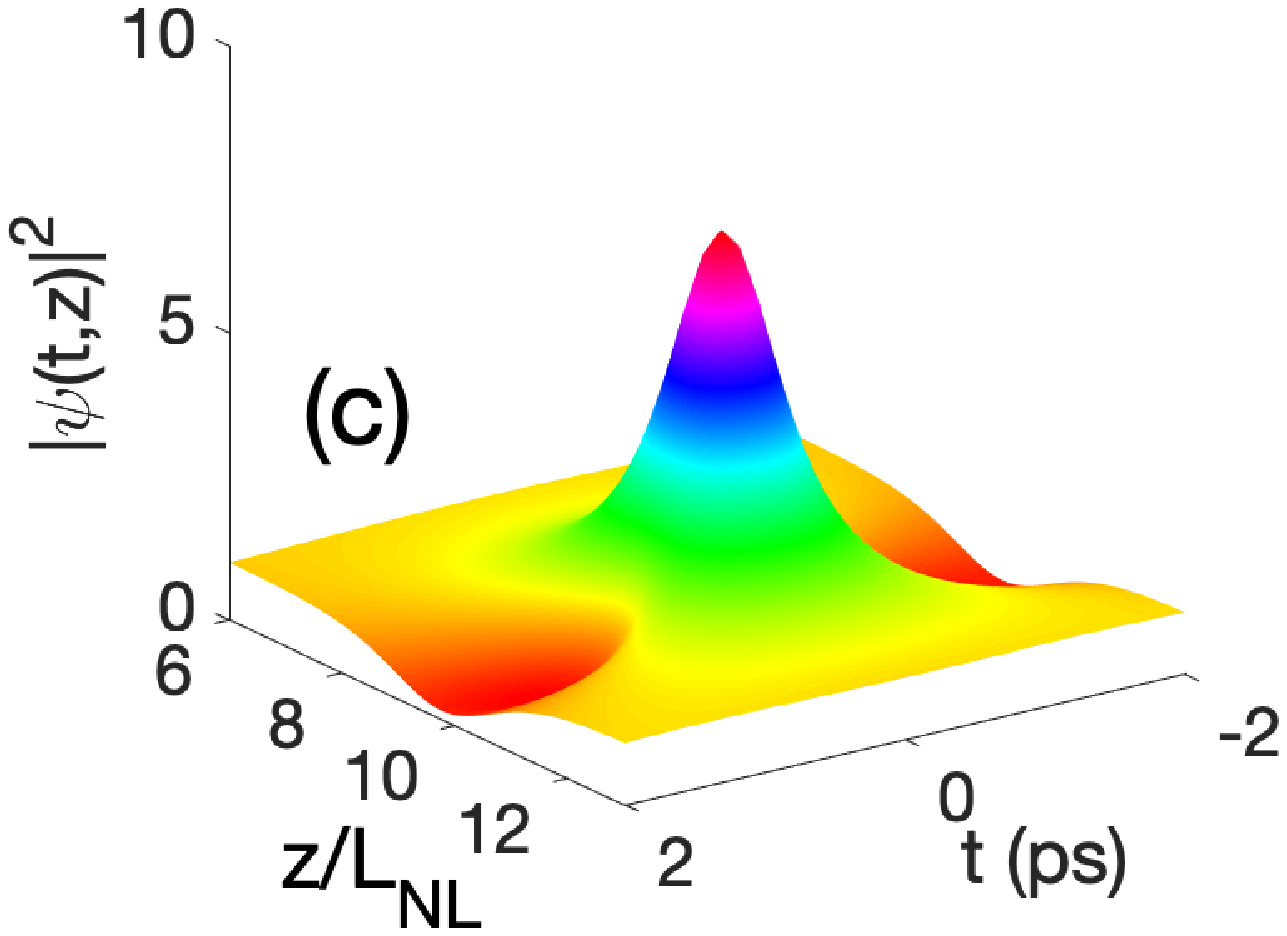}
\caption{Propagation of the perturbed plane wave intensity and wave pattern formation in presence of nonlocality for (aj)$_{j=1,2}$: $\beta_6=-0.1$ ps$^6$mm$^{-1}$, $\beta_4=-0.8$ ps$^4$mm$^{-1}$, $\beta_2=-1.5$ ps$^2$mm$^{-1}$ and (bj)$_{j=1,2}$:  $\beta_6=-0.5$ ps$^6$mm$^{-1}$, $\beta_4=-0.8$ ps$^4$mm$^{-1}$, $\beta_2=-1.5$ ps$^2$mm$^{-1}$, with $\gamma=0.01$,  $\Omega=0.75$ THz, $P_0=1$W, and $\alpha=0.5$ W$^{-1}$.mm$^{-1}$. The upper line shows the evolution of the perturbed plane wave, while the lower line displays the spectral amplitude versus the propagation distance $z$ and the frequency shift.}\label{fig18}
\end{figure}

The results of numerical simulations on the MI in the presence of the quadratic, quartic, and septic dispersions are presented in Figs.~\ref{fig16}-\ref{fig18} in the absence and presence of the nonlocality. To remind, the effect of the sextic dispersion is governed by the coefficient $\beta_6$ that can be positive or negative. In absence of the nonlocality, the MI is manifested by the patterns of Fig.~\ref{fig16}, where we have fixed $\beta_4=-0.8$ ps$^4$mm$^{-1}$, $\beta_2=-1.5$ ps$^2$mm$^{-1}$, while $|\beta_6|$ is attributed increasing values. The frequency domain shows some frequency shifts from $z=15$$L_{NL}$ that evolve as distance increases, showing a maintained instability of the CW for $\beta_6=-0.1$ ps$^6$mm$^{-1}$. In the same context, when $|\beta_6|$ is increased to $0.5$ ps$^6$mm$^{-1}$, a recombination phenomenon affects the temporal frequency of the RWs. In the same direction, the dynamics of MI shows multiple frequency shifts that testify to an early instability occurrence under sustained CW disintegration, as shown in Fig.~\ref{fig16}(b2). Fig.~\ref{fig17}(a) shows how the AB breather evolves during propagation, which illustrates the strong impact of propagation distance on the development of MI through which the multi-alternation behavior can be attributed to energy redistribution over the propagation distance.  In the same context and based on the AB patterns of Fig.~\ref{fig16}(bj)$_{j=1,2}$, the impact of reducing the magnitude of $|\beta_2|$ is shown in Fig.~\ref{fig17}(b) and supports the interplay between the quadratic and sextic dispersion in maintaining highly localized structures. It is obvious that decreasing $|\beta_2|$ contributes to reducing the amplitude of the rogue waves, which are indubitably of type A according to the available nomenclature. In the presence of the nonlocality, the features of instability given by Fig.~\ref{fig16} completely change as shown in Fig.~\ref{fig18}. In general, the inclusion of the nonlocality in the presence of the sextic dispersion affects the propagation of the instability. The nonlinear manifestation of MI is driven by AB lattice of type A for  $\beta_6=-0.1$ ps$^6$mm$^{-1}$, while the type-B breather patterns are obtained for $\beta_6=-0.5$ ps$^6$mm$^{-1}$. The propagation features are also different in terms of frequency shifts, where the distances between successive shifts get reduced with $|\beta_6|$ increasing. This seems to support efficient soliton fiber communication as it enhances regular bits of information release [see Fig.~\ref{fig18}(b2)].   Interestingly, each of the individual elements of the detected bit displays a Peregrine soliton-like shape [see Fig.~\ref{fig18}(c)] that is a straightforward signature of MI. Therefore, the MI approach is a way to make use of the quartic and sextic dispersions to balance higher-order nonlinear effects.  This results in the shape of the obtained quartic-sextic optical pulses being preserved while propagating along the optical fiber.

More recently, experimental and theoretical investigations of localized waves on plane-wave background have demonstrated that their dynamics is closely related to RWs, mainly including the AB~\cite{AkhmedievTMP1986}, Peregrine rogue wave~\cite{PeregrineJAMS1983}, and Kuznetsov-Ma breather~\cite{Kuznetsov1977}. RWs are relatively large and spontaneous waves whose appearance may result in catastrophic damage~\cite{DraperMObs1965,SolliNat2007,KharifBook2009}. It is known that the generation of RWs is closely related to modulation instability (MI), which is a complex nonlinear process exhibiting emergent behavior and strong sensitivity to initial conditions~\cite{OnoratoPR2013,DudleyNatPhot2014,DudleyOptExp2009,AkhmedievJOpt2016}. The MI is frequently studied within the NLS equation, which models the evolution of weakly nonlinear dispersive wave packets. One can therefore study MI's initial (i.e., linear) stage by linearizing the NLS equation around the CW background.   It is a well-known fact that the linear stage of the MI is characterized by the exponential growth of all the perturbations falling in the region of the Fourier spectrum below a certain threshold through the amplification of sideband frequencies and is closely associated with the concept of self-localized waves, or solitons. The linearization ceases to be valid when the amplitude of the growing perturbation becomes comparable to the CW background, that is, at the nonlinear stage of MI, corresponding to the long-term spatiotemporal dynamics. In this respect, a particular scenario of the MI development strongly depends on the type of initial conditions considered, namely noisy perturbations of the plane wave, localized perturbations of the plane wave, and harmonic perturbations of the plane wave, respectively. For an initial condition consisting of a CW field perturbed by a sine-modulation at the frequency of maximum MI gain, the dynamics exhibits a recurrent behavior where AB-like structures are formed periodically. In this context, the nonlinear stage of the MI is described by an exact solution of the NLS, known as AB~\cite{AkhmedievJETP1985,AkhmedievMP1986,AkhmedievTMP1987,AkhmedievJETP1988,AkhmedievPLA2009,EstellePRE2015,EstellePRE2016,EstellePRE2018}. For noisy initial conditions, structures close to the AB, Kuznetsov-Ma solitons, and Peregrine solitons, which are breather solutions of the 1D NLS equation, emerge locally from the propagation of a perturbed modulational unstable CW background~\cite{OnoratoPR2013,DudleyNatPhot2014,DudleyOptExp2009,AkhmedievJOpt2016}. In fact,  the simplest first-order solutions that are either periodic in space and localized in time or periodic in time and localized in space are referred to as Kuznetsov-Ma breathers and AB, respectively. Taking both solutions' periods towards infinity makes it possible to approach the limit of the doubly localized Peregrine breather.

The Darboux transformation method has explored interaction properties between rogue waves and breathers~\cite{LiCSF2022}. The correspondence relation between MI and types of fundamental nonlinear excitation in optical fibers with both TOD and FOD terms have been reported, including mainly the AB, RW, Kuznetsov-Ma breather, periodic wave, W-shaped soliton train, rational W-shaped soliton, anti-dark soliton, and nonrational W-shaped soliton~\cite{DuanPRE2017}. It has been shown that the dynamical process of Kuznetsov-Ma breather involves at least two distinctive mechanisms, where the MI plays dominant roles in the mechanism of Kuznetsov-Ma breather admitting weak perturbations, while the interference effect plays a dominant role for the Kuznetsov-Ma breather admitting strong perturbations~\cite{ZhaoPRE2018}. Predictable rogue wave solutions of the (2 +1)-dimensional B-type Kadomtsev-Petviashvili equation by using symbolic calculation methods~\cite{PengPLA2018}. It has been reported that different MI gain distribution can induce different rogue wave excitation patterns in a two-mode nonlinear fiber whose dynamics is described by two-component coupled NLS equations~\cite{ZhaoPRE2014}.

\section{Concluding remarks}\label{s5}

Based on recent experimental findings, the importance of wave propagation in Kerr media in the presence of quadratic, quartic, and sextic dispersions has gained significant interest. While some orientations pay attention to finding exact soliton solutions to such models, we have devoted the present work to the study of MI in presence of weakly nonlocal nonlinearity. In this context, considering the nonlocal response as a perturbation to the Kerr nonlinearity, we have evaluated the development of MI under different contexts of dispersions both analytically and numerically. In order to find an expression for the MI gain, the linear stability analysis of a CW  has been used, from which we have further derived the optimum modulation frequency. Using the latter, the salient effect of the quartic and the sextic dispersions, and their competition against the nonlocal nonlinearity,  on the MI spectra have been comprehensively addressed.   In many cases, the dynamics of systems exhibiting MI is governed to leading order by the one-dimensional cubic NLS equation. Indeed, as noticed in this work, the NLS equation admits a plane wave solution whose condition of instability has been widely discussed in the literature. When the plane wave is unstable, its disintegration into solitonic structures takes place. Fortunately, the condition for instability, also known as the {\it Benjamin-Feir} criterion, gives rise to regions of parameters where bright-type soliton (localized envelope) exact solution exists with a vanishing amplitude at $|x|\rightarrow\infty$. With rare exceptions~\cite{AkhmedievSpringe2005}, dissipative systems are non-integrable, with a good example being the cubic complex Ginzburg-Landau (CGL) equation, which also admits a plane wave solution. The latter becomes unstable under the Lange-Newell criterion, a generalized version of the Benjamin-Feir condition, due to the competing effects of gains and losses~\cite{PelapJPSJ2001,KenfackJPSJ2003}. Under such a condition, the plane wave also breaks up into solitonic structures, albeit the CGL equation is non-integrable. This also explains why, although nonintegrable, the pure quartic NLS equation has been found to support areas of instability under the strong action of nonlocality, both for negative and positive values of $|\beta_4|$.

For OMF involving higher-order polynomials, the spectrum of MI has been analyzed numerically,  showing the strong input of the nonlocality in the emergence of MI sidebands when the sextic dispersion is involved, under the strong effect of the quartic and quadratic dispersions. Using such predictions, the MI gain has been comprehensively exploited in finding regions of parameters where MI could take place. The calculations revealed the existence of extended tails of the MI gain due to the nonlocality and well-balanced higher-order even dispersions. To confirm such analytical predictions,  direct numerical simulations have been carried out on the generic model, where investigations have been based on various combinations of the dispersions terms, the nonlinearity, and the nonlocality. Under pure quartic conditions ($\beta_2=0$ and $\beta_6=0$) the CW was stable in the absence of nonlocality, while the latter ensured the appearance of nonlinear structures with solitonic form, showing locally and periodically AB-like evolutions. The joint presence of the quadratic and quartic terms was found to support CW disintegration, but including the nonlocal terms brought about extended sneaky breathers that were spatially more extended with a broader frequency shift. However, strong quartic dispersion brought about early MI development but the frequency shift was too extended as a manifestation of erratic patterns, under strong output signal intensity. With the right choice of $\beta_2$ and $\beta_4$, it was also possible to generate extended frequency shift modulation mediated by single- and double-humped breathers, with strong input from nonlocal nonlinearity. More interestingly, the presence of the sextic dispersion in the previous context brought about other features of instability, among which the possibility of AB generated via the activation of MI, with exotic behaviors highly dependent on the nonlocality. For example, one could generate highly localized repetitive impulses by well-balancing the sextic dispersion and the nonlocality term.

Solitons or other nonlinear objects are of relevant importance in physics in general since, as mathematical models, they provide a better understanding of physical models, and they may lead to physical applications.  So, finding exact solutions of the sixth-order dispersive NLS equation with a weak nonlocal nonlinearity is a task of relative importance. For example, exact solutions may {\bf (i)} help to choose appropriate experimental parameters, {\bf (ii)} provide a way for probing the validity of the nonlinear evolution equations, {\bf (iii)} help to analyze the stability of solutions, {\bf (iv)} check the numerical analysis of the nonlinear evolution equations, and {\bf (v)} help to explain the formation and the propagation of different kinds of patterns in the system under consideration as well as their long-time evolution. However, beyond the limitations inherent to exact methods for the derivation of exact soliton solutions, MI can produce a soliton train formation with 26 solitons formed in an elongated attractive Bose-Einstein condensate in an axisymmetric harmonic trap~\cite{CarrPRL2004}, which is a groundbreaking result. On the contrary, using the inverse scattering methods or the Hirota bilinear method to generate 26 solitons is quite cumbersome, if not possible. Additionally,  the spontaneous development of MI seeded from noise has also been shown to underlie the initial stages of fiber supercontinuum generation seeded by continuous wave (CW) radiation or picosecond or nanosecond pulses, which remains a challenging ambition of exact solution seekers. 

\section*{Acknowledgements}
CBT thanks the Kavli Institute for Theoretical Physics (KITP), University of California Santa Barbara (USA), where this work was supported in part by the National Science Foundation Grant no.{\bf NSF PHY-1748958}, NIH Grant no.{\bf R25GM067110}, and the Gordon and Betty Moore Foundation Grant no.{\bf 2919.01}.

\end{document}